\documentclass[nonacm, sigconf, screen, authorversion]{acmart}
\usepackage{amsmath,amsfonts}
\usepackage{algorithmic}
\usepackage{graphicx}
\usepackage{textcomp}
\usepackage{xcolor}
\usepackage{lmodern}
\def\BibTeX{{\rm B\kern-.05em{\sc i\kern-.025em b}\kern-.08em
    T\kern-.1667em\lower.7ex\hbox{E}\kern-.125emX}}

\usepackage{paralist}
\usepackage{multirow}

\usepackage{xspace}

\usepackage{hyperref}
\AtBeginDocument{%

}
\usepackage[framemethod=TikZ]{mdframed}

\usepackage{listings}

\usepackage[shortlabels,inline]{enumitem}\setitemize{noitemsep,topsep=0pt,parsep=0pt,partopsep=0pt,leftmargin=*}

\usepackage{subcaption}

\usepackage{csquotes}
\usepackage{booktabs}
\usepackage{tabularx}

\usepackage{ifdraft}
\usepackage{ifthen}
\newcommand{\add}[1]{{#1}}

\newboolean{debug}
\setboolean{debug}{false}

\ifthenelse{\boolean{debug}}{%
  \usepackage{showframe}                 
  \usepackage{todonotes}                 
  \newcommand{\xnote}[1]{\todo[inline]{#1}}
  \newcommand{\xynote}[2]{\todo[inline]{#1: #2}}
  
  \usepackage{hyperref}                  
  \pagestyle{plain}                      
  \newcommand{\sosym}[1]{{#1}}
  \newcommand{\revone}[1]{{#1}}
  \newcommand{\revtwo}[1]{{\color{blue}#1}}

  \newcommand{\hiddenparagraph}[1]{\paragraph{#1}}
  \newcommand{\smallnote}[1]{
    \par\noindent\makebox[\textwidth][c]{%
      \fbox{
        \begin{minipage}{0.9\textwidth}
          \scriptsize{\color{gray}{#1}}
        \end{minipage}
      }
    }
  }
  
}{%
  \usepackage[disable]{todonotes}        
  \newcommand{\xnote}[1]{}
  \newcommand{\xynote}[2]{}
  
  \hypersetup{hidelinks}                 
  \pagestyle{empty}                      
  \newcommand{\sosym}[1]{{#1}}
  \newcommand{\revone}[1]{{#1}}
  \newcommand{\revtwo}[1]{{#1}}

  \newcommand{\hiddenparagraph}[1]{}
  \newcommand{\smallnote}[1]{}
  
}

\newcommand{\journal}[1]{}               

\newcommand{\cps}{vehicle\xspace}
\newcommand{\cpss}{vehicles\xspace}

\newcommand*{\ie}{i.e.,\@\xspace}
\newcommand*{\eg}{e.g.,\@\xspace}
\newcommand*{\cf}{cf.\@\xspace}

\newcounter{requirement}[section]

\newcounter{derivationrulescounter}
\definecolor{keyword}{RGB}{154,0,154}
\definecolor{string}{RGB}{40,90,0}
\definecolor{commentcolor}{RGB}{110,55,0}
\definecolor{stringBlue}{rgb}{0.58,0,0.82}
\definecolor{tfvarcolor}{RGB}{0,0,204}

\lstdefinelanguage{tf}[]{}{
  stringstyle=\ttfamily\color{string},
  morestring=[b]",
  keywordstyle=\bfseries\color{tfvarcolor},
  keywordstyle=[3]\bfseries\color{keyword},
  alsoletter={\_},
  keywordsprefix=\$,
  keywords=[3]{not, assign, do, where, folding, list, opt, :-},
  morecomment=[l][\itshape\color{commentcolor}]{//},
  morecomment=[s][\itshape\color{commentcolor}]{/**}{*/}
}

\mdfdefinestyle{rqanswer}{backgroundcolor=gray!10,roundcorner=3pt,outerlinewidth=1,
innerleftmargin=3pt,innerrightmargin=3pt}

\lstdefinelanguage{ma}[]{Java}{
  morekeywords={component, port, in, out, inv, package, import, 
  connect, 
  autoconnect}
}

\usepackage[nomain,acronym,nohypertypes={acronym}]{glossaries}

\newcommand{\singular}[1]{\gls{#1}\xspace}
\newcommand{\plural}[1]{\glspl{#1}\xspace}
\newcommand{\abkuerzung}[2]{\newacronym{#1}{#1}{#2}}

\abkuerzung{AADL}{Architecture Analysis \& Design Language}

\newacronym{CC}{C\&C}{component and connector}

\abkuerzung{CD}{Class Diagram}

\abkuerzung{CBSE}{Component-Based Software Engineering}

\abkuerzung{CPPS}{Cyber-Physical Production System}

\abkuerzung{CFG}{Context-Free Grammar}

\abkuerzung{DSL}{Domain-Specific Language}

\abkuerzung{DT}{Digital Twin}
\newcommand{\dt}{\singular{DT}}
\newcommand{\dts}{\plural{DT}}

\abkuerzung{FSM}{Finite-State Machine}

\abkuerzung{GPL}{General-Purpose Programming Language}

\abkuerzung{LPL}{Language Product Line}

\abkuerzung{MDD}{model-driven development}

\abkuerzung{MDSE}{Model-Driven Systems Engineering}

\abkuerzung{MBSE}{model-based systems engineering}

\abkuerzung{M2M}{model-to-model}

\abkuerzung{M2T}{Model-to-Text}

\abkuerzung{OCL}{Object Constraint Language}

\abkuerzung{PaaS}{platform as a service}

\abkuerzung{SC}{statechart}

\abkuerzung{SLE}{Software Language Engineering}

\abkuerzung{SPL}{Software Product Line}

\abkuerzung{SysML}{Systems Modeling Language}

\abkuerzung{UML}{Unified Modeling Language}

\usepackage{pgf}
\usepackage{tikz}
\usepackage{pgfplots}
\pgfplotsset{compat=1.18}
\usepackage{pgfplotstable}
\usepgfplotslibrary{colorbrewer}
\usetikzlibrary{shapes,backgrounds,mindmap,trees,calc}
\usepackage{colortbl}

\colorlet{circuit}{red}
\colorlet{cad}{blue}
\colorlet{eea}{orange}
\colorlet{code}{violet}
\colorlet{data}{magenta}
\colorlet{math}{teal}
\colorlet{fem}{gray}
\colorlet{simulation}{cyan}
\colorlet{workflow}{olive}
\colorlet{process}{purple}
\colorlet{variability}{green!25!black}
\definecolor{heatmaphot}{HTML}{c96a52}
\definecolor{heatmapcold}{HTML}{568da7}

\pgfplotscreateplotcyclelist{modelkinds}{
    {fill=circuit},
    {fill=cad},
    {fill=eea},
    {fill=code},
    {fill=data},
    {fill=math},
    {fill=fem},
    {fill=simulation},
    {fill=workflow},
    {fill=process},
    {fill=variability}}

\newcommand{\boxplot}[7][blue]{%
	\filldraw[fill=#1!30,draw=#1,line width=.2mm] let \n{boxxl}={#2-0.2}, \n{boxxr}={#2+0.2} in (axis cs:\n{boxxl},#4) rectangle (axis cs:\n{boxxr},#5);   
	\draw[line width=.2mm] (axis cs:#2,#5) -- (axis cs:#2,#7);                                                                           
	\draw[line width=.2mm] let \n{whiskerl}={#2-0.15}, \n{whiskerr}={#2+0.15} in (axis cs:\n{whiskerl},#7) -- (axis cs:\n{whiskerr},#7);        
	\draw[line width=.2mm] (axis cs:#2,#4) -- (axis cs:#2,#6);                                                                           
	\draw[line width=.2mm] let \n{whiskerl}={#2-0.15}, \n{whiskerr}={#2+0.15} in (axis cs:\n{whiskerl},#6) -- (axis cs:\n{whiskerr},#6);        

	\draw[line width=.5mm, color=#1] let \n{boxxl}={#2-0.2}, \n{boxxr}={#2+0.2} in (axis cs:\n{boxxl},#3) -- (axis cs:\n{boxxr},#3);             
}

\pgfplotsset{
    legend image with color/.style={
        legend image code/.code={%
            \node[rectangle, fill=#1, anchor=center] at (0pt,0.5pt) { };
        }
    },
}

\begin{document}

\setcopyright{none}
\acmConference[MODELS '24]{ACM/IEEE 27th International Conference on Model Driven Engineering Languages and Systems}{September 22--27, 2024}{Linz, Austria}
\acmBooktitle{ACM/IEEE 27th International Conference on Model Driven Engineering Languages and Systems (MODELS '24), September 22--27, 2024, Linz, Austria}
\acmDOI{10.1145/3640310.3674100}
\acmISBN{979-8-4007-0504-5/                             24/09}

\title[Modeling Languages for Automotive Digital Twins]{A Survey of Modeling Languages for Automotive Digital Twins}

\settopmatter{authorsperrow=3}

\author[Pfeiffer]{Jérôme Pfeiffer}
\affiliation{%
  \institution{University of Stuttgart}
  \city{Stuttgart}
  \country{Germany}}
\email{jerome.pfeiffer@isw.uni-stuttgart.de}
\orcid{0000-0002-8953-1064}

\author[Fuchß]{Dominik Fuchß}
\affiliation{%
  \institution{Karlsruhe Institute of Technology }
  \city{Karlsruhe}
  \country{Germany}}
\email{dominik.fuchss@kit.edu}
\orcid{0000-0001-6410-6769}

\author[Kühn]{Thomas Kühn}
\affiliation{%
  \institution{Martin Luther University Halle-Wittenberg}
  \city{Halle}
  \country{Germany}}
\email{thomas.kuehn@informatik.uni-halle.de}
\orcid{0000-0001-7312-2891}


\author[Neumann]{Dirk Neumann}
\affiliation{%
  \institution{Karlsruhe Institute of Technology }
  \city{Karlsruhe}
  \country{Germany}}
\email{dirk.neumann@kit.edu}
\orcid{0000-0003-1999-639X}

\author[Neimöck]{Christer Neimöck}
\affiliation{%
  \institution{T-Systems GmbH}
  \city{Frankfurt}
  \country{Germany}}
\email{C.Neimoeck@t-systems.com}
\orcid{0009-0000-4340-0440}


\author[Koziolek]{Anne Koziolek}
\affiliation{%
  \institution{Karlsruhe Institute of Technology }
  \city{Karlsruhe}
  \country{Germany}}
\email{koziolek@kit.edu}
\orcid{0000-0002-1593-3394}

\author[Wortmann]{Andreas Wortmann}
\affiliation{%
\institution{University of Stuttgart}
  \city{Stuttgart}
  \country{Germany}}
\email{wortmann@isw.uni-stuttgart.de}
\orcid{0000-0003-3534-253X}

\begin{abstract}
%
The demand for digital twins and suitable modeling techniques in the automotive industry is increasing rapidly. 
%
Yet, there is no common understanding of digital twins in automotive, nor are there modeling techniques established to create automotive digital twins. 
%
Recent studies on digital twins focus on the analysis of the literature on digital twins for automotive or in general and, thus, neglect the industrial perspective of automotive practitioners. 
%
To mitigate this gap between scientific literature and the industrial perspective, we conducted a questionnaire survey among experts in the German automotive industry to identify
\begin{inparaenum}[(i)]
\item the desired purposes for and capabilities of digital twins,
\item the modeling techniques related to engineering and operating digital twins across the phases of automotive development, and
\item \sosym{the role informal models play during automotive development.}
\end{inparaenum}
%
%
%
To this end, we contacted 189 members of the Software-Defined Car research project and received 96 responses. 
The results show that digital twins are considered most useful in the usage and support phase of automotive development, representing vehicles as-operated. Moreover, simulation models, source code, and business process models are currently considered the most important models to be integrated into a digital twin alongside the associated, established tools.
\sosym{Furthermore, informal models are frequently created digitally, e.g., using PowerPoint\texttrademark, mainly employed for communication and documentation during the conception and development phase.}

\end{abstract}


\maketitle

\section{Introduction}
\label{sec:intro}

\dts~\cite{kritzinger2018digital,tao2018digital} are complex software systems that connect to an actual system, \eg a (simulated) vehicle, obtain data from it, represent it based on this data, and use this data to optimize its behavior~\cite{tc1842020iso,dtcCapabilitiesTable,eramo2021conceptualizing}.
They are used in a variety of domains, including agriculture~\cite{DBLP:conf/ghtc/AlvesSMTKSAL19}, manufacturing~\cite{BBD+21b}, medicine~\cite{DBLP:journals/cmpb/LauzeralBKGRHC19}, and many more~\cite{DJR+22}.
Typical purposes of digital twins include monitoring~\cite{DBLP:conf/modelica/MagargleJMDKKBP17}, validating~\cite{hua2022}, predicting~\cite{rajesh2019digital}, and optimizing the behavior of the actual system.
For this, \dts require different properties~\cite{dtcCapabilitiesTable}, such as, collecting data, model processing, simulation, bi-directional communication. 
Related studies collected various definitions of \dts across domains from the literature~\cite{DJR+22} and investigated how \dts can be engineered for the production of automotive and what their added-value could be~\cite{biesinger2019survey}. 
While other related studies focusing on modeling languages~\cite{WBCW20,ABH+21} and model-based engineering~\cite{DBLP:journals/sosym/LiebelMTLH18} exist, they either only query existing literature or are tailored to a certain domain within the automotive development life cycle, \eg to embedded systems~\cite{DBLP:journals/sosym/LiebelMTLH18,DBLP:journals/jsa/AkdurGD18}. 
Concerning informal models, different studies have examined the use of informal models~\cite{storrle_how_2017}, sketches~\cite{baltes_sketches_2014, walny_follow_2011}, and diagrams~\cite{cherubini_lets_2007} in software development. However, no study focusing on the utilization of informal model in the automotive development exists.

At the center of digital twins are formal models, representing the actual system, and data, obtained from it, to which the models give meaning. 
Hence, digital twins are model-centric software systems and modeling techniques shape the capabilities and properties digital twins can express and reason about.
This is also reflected by large software vendors, such as Microsoft or Amazon, as well as the Eclipse Foundation that actively develop dedicated modeling languages for \dts~\cite{LPT+22,PLW+22}.

In addition, informal models are already widely used in software development \cite{5}, as they ease the communication and documentation of software. 
Furthermore, they support multiple software development phases \cite{baltes_sketches_2014,10}. 
Consequently, research on approaches to deal with informal diagrams and sketches often focus on software development \cite{6,9,11,13}. 
Notably though, the use of informal models for the development of CPSs or the development with a digital twin is rarely studied.
\revone{
A future automotive digital twin could not only include formal engineering models but also the rich ecosystem of informal models—sketches, diagrams, slides, notes, and conceptual drawings—by automatically interpreting, structuring, and transforming them into actionable, machine-readable knowledge. 
Consequently informal models would become a living input source that continuously enriches the twin throughout the vehicle’s lifecycle.
}

A study surveying automotive industry experts on the expectations, use, and formal, as well as, informal modeling techniques for \dts in their domain is missing.

To remedy this, we investigate the purposes, expected properties, \sosym{formal, and informal} modeling techniques for \dts in the context of the German research project Software-Defined Car (SofDCar).\footnote{SofDCar website: \url{https://sofdcar.de/language/en/}.}
While SofDCar features German partners only, its consortium includes (i) worldwide leading automotive companies and suppliers, such as Bosch, Mercedes-Benz, ZF Friedrichshafen; (ii) renown software companies working in automotive, including T-Systems International and Vector Informatik, as well as (iii) leading research institutes on software engineering, such as Karlsruhe Institute of Technology, the University of Stuttgart, and the FZI Research Center for Information Technology.
With many of the consortium participants contributing to the shape of digital twins in automotive internationally, we expect the insights from this survey to generalize beyond Germany. 

\revtwo{
To gather insights from the SofDCar experts, we designed and administered a questionnaire. 
Because we assume that digital twins are still uncommon in automotive practice, the first part focuses on experts’ understanding of and expectations for digital twins. 
This also sets the context for digital twins.}
The second part examines current modeling practices, including the use of modeling languages, tools, and informal models. This allows us to identify existing models that could potentially support future digital twin development.
Based on the experts’ responses, we address the following research questions (RQs):

\vspace{5pt}
\begin{mdframed}[style=rqanswer]
\revone{
\textbf{RQ:} When developing digital twins for the automotive industry, what are the expectations of digital twins, and which modeling languages, tools, and practices should be taken into account for future automotive digital twins?
}
\begin{enumerate}
\item[\textbf{RQ1:}] How is the \dt understood in the automotive industry? 
    \begin{enumerate}
    \item[\textbf{RQ1.1:}] For which phases of automotive development are \dts important?
    \item[\textbf{RQ1.2:}] What are desired properties of \dts?
    \item[\textbf{RQ1.3:}] What are desired purposes of using \dts?
    \item[\textbf{RQ1.4:}] How do these purposes change in relation to different phases of automotive development?
    \end{enumerate}
\item[\textbf{RQ2:}] Which modeling languages and modeling tools are currently employed in the automotive industry?
\begin{enumerate}
        \item[\textbf{RQ2.1:}] Which kinds of models are important during automotive development?
        \item[\textbf{RQ2.2:}] How important are which models in the phases of automotive development?
        \item[\textbf{RQ2.3:}] Which tools are used to create and maintain these models?    
    \end{enumerate}

\item[\textbf{RQ3:}] How pervasive are informal models in the automotive industry?
\begin{enumerate}
\item[\textbf{RQ3.1:}] For which purpose are informal models used?
\item[\textbf{RQ3.2:}] For which phases of automotive development are informal models employed?
\item[\textbf{RQ3.3:}] How are informal models persisted?
\item[\textbf{RQ3.4:}] Which kinds of models are informally designed?
\item[\textbf{RQ3.5:}] How are informal models re-/used?
\end{enumerate}

\end{enumerate}
\end{mdframed}

The results of our survey indicate that \dts are especially important when monitoring and collecting data from the vehicle in its operation. 
Furthermore, at least one kind of modeling language is currently employed by all participants, indicating that using these models in the context of digital twins will be the key to their successful deployment in practice. 
\sosym{Last but not least, most participants employed informal models during the conception and development phase to communicate and document design decisions. Surprisingly, most participants frequently create informal models digitally, e.g., using PowerPoint.}

\revone{
This extended study has the following lessons learned: 
\begin{itemize}
    \item Model integration is crucial, with mathematical/physical and simulation models seen as essential.
    \item Automotive \dt modeling is domain-specific, with surprisingly low use of UML/SysML compared to general \dt literature, and strong dependence on specialized tooling. 
    \item Tool integration is a major challenge, indicating the need for research on connecting diverse models and tools (\eg, via knowledge graphs, APIs, exchange formats). 
    \item Informal models remain common for communication, creating opportunities for AI to convert them into formal models and support intelligent modeling assistants in future workflows.
\end{itemize}
}

\revone{
This paper is based on a conference paper~\cite{DBLP:conf/models/PfeifferF0L0NSK24} and a preliminary report on informal models~\cite{fuchss_expert_2023}, that we combined, as in the initial conference paper we omitted the inclusion of the inquired use of informal models. 
In this extension we include the results and the analysis of the third part of our questionnaire about informal models. 
This part about informal was handed out to the automotive experts from the SofDCar consortium together with the questions about \dts and formal models. 
Extending our study with this part entails a new research question and two sections in the results and discussion section that present and analyze our findings. 
}

In the remainder, \autoref{sec:background} introduces background information before \autoref{sec:methodology} explains the design of our questionnaire survey, and \autoref{sec:results} presents its results.
Afterward, \autoref{sec:discussion} discusses the results as well as threats to validity, while \autoref{sec:relatedwork} discusses related studies.
Finally, \autoref{sec:conclusion} concludes this paper.


\section{Background}
\label{sec:background}

Before diving into the conducted study, in this section, we give a brief introduction to the notion of Digital Twins as well as to the diagrams we employ, i.e., box plots, heatmaps, and Sankey diagrams.
The knowledgeable reader can safely skip this section.

\subsection{Formal and Informal Modeling}

\subsection{Digital Twins}
\label{sec:background_dt}
\sosym{Henceforth, we define a digital twin (\dt) as: 
\emph{\enquote{a software comprising models and services to purposefully represent and manipulate its original system across its complete lifecycle}}~\cite{kirchhof2020model,BDH+20,kritzinger2018digital}.
This definition acknowledges that a \dt mirrors an original system, providing means to digitally represent that system, and to manipulate it. 
Moreover, it highlights that the models and services of a \dt fulfill a specific purpose, which governs the models' abstraction and pragmatism.}
Furthermore, twinning means that changes to the original system are automatically reflected in the \dt and changes in the \dt are automatically reflected in the original system.
To this end, \dts must establish a representation of their counterpart, along with a mechanism for transmitting changes between both systems~\cite{9626349}.
Based on this, \dts enable a variety of value-adding services, such as, monitoring, behavior prediction~\cite{252}, process optimization~\cite{BHK+21b}, predictive maintenance, and many more~\cite{tao2019digital, 9438560}. 
As parts of our survey are based on our definition and experience with digital twins, the last question in our survey requested the agreement of the participants with our definition. 
We discuss the results to this answer in \autoref{sec:discussion}. 

\subsection{Diagrams}
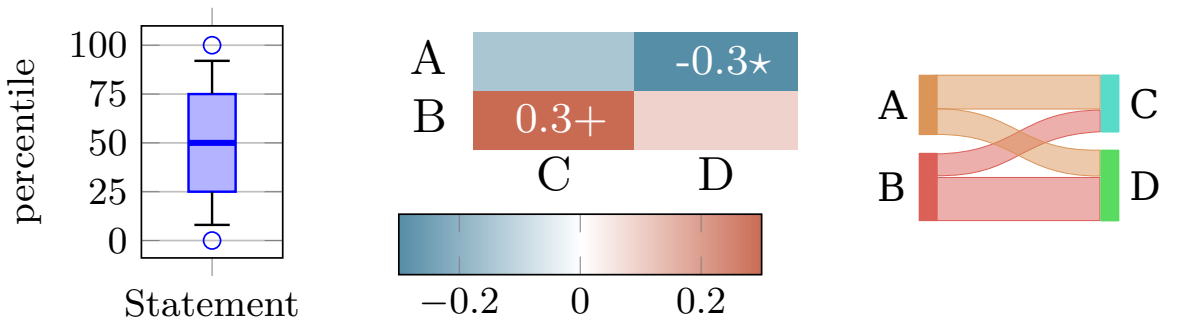
\begin{figure}
    \begin{subfigure}{0.32\linewidth}
    \centering
    \begin{tikzpicture}[scale=.8]
	\begin{axis}[width=2.75cm, height=3.5cm, xmin=-0.5, xmax=0.5,
		   ymin=1, ymax=100,
		   ybar, ymajorgrids, enlargelimits=0.1,
		   legend style={anchor=north east,legend columns=6},
		   scaled ticks=false, tick label style={/pgf/number format/fixed},
		   xtick={0}, 
          xticklabels={Statement},
          x tick label style={rotate=0,anchor=north},
          ytick={0,25,50,75,100},
          ylabel={percentile}
		  ]

    \boxplot[blue]{0}{50}{25}{75}{8}{92}
    \addplot[only marks,mark=o,fill=none,blue,forget plot] coordinates {(0,0) (0,100)};
    
	\end{axis}
\end{tikzpicture}
    \subcaption{Box plot}
    \label{fig:boxplot-example}
    \end{subfigure}
    \begin{subfigure}{0.32\linewidth}
    \centering
        \begin{tabular}{rcc}
    A &\small  \cellcolor{heatmapcold!50!white} \phantom{M} &\small \cellcolor{heatmapcold} \textcolor{white}{-$0.3\star$}\\
    B &\small  \cellcolor{heatmaphot} \textcolor{white}{$0.3$+}&\small \cellcolor{heatmaphot!30!white} \phantom{M}\\
      & C & D
    \end{tabular}
    
    \begin{tikzpicture}[scale=.8]
    \pgfplotscolorbardrawstandalone[
        colormap={my colormap}{
            color(0cm)=(heatmapcold); color(1.5cm)=(white); color(3.0cm)=(heatmaphot);
        },
        height=2cm,
        point meta min=-0.3,
        point meta max=0.3,
        parent axis width/.initial=3.0cm,
        parent axis height/.initial=1cm,
        colorbar horizontal,
        ] 
    \end{tikzpicture}
    \subcaption{Heatmap}
    \label{fig:heatmap-example}
    \end{subfigure}
    \begin{subfigure}{0.32\linewidth}
    \centering
    \raisebox{2em}{\includegraphics[width=2.0cm]{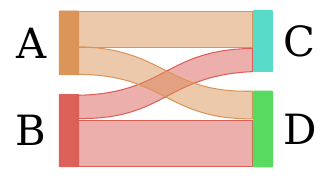}}
    \subcaption{Sankey diagram}
    \label{fig:sankey-example}
    \end{subfigure}
    \caption{Illustrative examples of the employed diagrams in our study.}
    \label{fig:example-diagrams}
\end{figure}

For the presentation of our results, we utilize a variety of diagrams.
Besides histograms and bubble charts, we employ box plots, heatmaps, and Sankey diagrams to visualize the information we gathered.
As the latter might be unfamiliar to some readers, \autoref{fig:example-diagrams} illustrates each of them.
First, box plots (\autoref{fig:boxplot-example}) visualize the distribution of answers given, especially for the four-point Likert scales.
On the x-axis, we show the different properties or statements the participants could rate, and on the y-axis, we show the rating of the participants.
The box indicates the range from 75th to 25th percentile, whereas the thick line indicates the median.
The whiskers represent the highest and lowest answer in the interquartile range, i.e., $1.5$ times the 75th minus the 25th percentile.
Any answer that is beyond the interquartile range from the median is considered an outlier, indicated by a circle.
Second, heatmaps (\autoref{fig:heatmap-example}) visualize the correlation between different aspects of a dataset.
Here, we show the correlation between different participants answering A or B in the first question and C or D in the second question.
\revone{The x and y-axis show the different options, whereas the cell color indicates the Pearson correlation coefficient between the two options. A legend indicates which color corresponds to which coefficient. In addition, we perform a T-test analysis to determine the significance of correlations. In case of a significant correlation the coefficient and a strength indicator is shown, the strong significance ($p < 0.05$) of the correlation between B and C is signified with a star in the cell, whereas the weak significance ($p < 0.1$) between A and D with a plus.}
Last but not least, Sankey diagrams (\autoref{fig:sankey-example}) visualize participants giving the same answers as a flow from one question to another.
Consequently, the width of the flow indicates the number of participants giving the same answers.
In \autoref{fig:sankey-example}, the flow from option A to option C highlights that most participants who chose A also choose C in the second question, whereas only a few opted for D.
\section{Methodology}\label{sec:methodology}
The creation of the questionnaire started with the selection of our research questions to be answered. 
They are based on the research gap in the current literature where no study exists that surveys experts from the automotive industry on their understanding, expectations, and employed modeling techniques for \dts in automotive development. 
Next, we created the initial version of the questionnaire based on previous work on \dts~\cite{BBD+21b,PLW+22,DHM+22,DJR+22,FHM+23} and model-driven software engineering~\cite{WBCW20,DBLP:journals/jot/GotzFRK20,DBLP:conf/sose/SchindewolfWKGS23}, and discussions with industry partners within the SofDCar project. 
After that, we conducted two preliminary studies, to end up with our final questionnaire comprising 54 questions. 
\revone{This questionnaire included the questions about the understanding of digital twins, the employed modeling languages and tools as well as the leveraged informal models. This questionnaire was handed out as one coherent document to the recipients.}
The average time required to fill out the survey is 15 minutes. 
Since all members of the project are German, the survey language is German, too. 
A PDF export of our questionnaire in German as well as all answers 
can be found online.\footnote{\label{reppack}Replication packages: \url{https://zenodo.org/doi/10.5281/zenodo.12529996} and \url{https://doi.org/10.5281/zenodo.8199850 }}
We anonymized the data to tackle re-identification risk. 
We created the survey with the tool LimeSurvey\footnote{Tool for the questionnaire: \url{https://www.limesurvey.org/}}.
LimeSurvey offers a CSV export that facilitates the import into other tools for data analysis. 
In particular, we utilized the \textit{Statistical Package for the Social Sciences} (SPSS)\footnote{Tool for analysis of participant's answers: \url{https://www.ibm.com/spss}} and custom Python scripts, for instance, to create the Sankey diagrams.
\revone{We performed a correlation analysis using the Pearson correlation coefficient and each time two-sided T-test.}

\subsection{Preliminary Studies}
We performed two preliminary studies before arriving at our final version of the questionnaire. 
We handed out the first draft of our questionnaire to institute members and collected feedback. 
The results showed that open-ended questions for both of our research questions hamper the analyzability of the responses. 
Therefore, we reworked the questionnaire and introduced Likert scales that ensure better analyzability. 
In the second preliminary study, we handed out the reworked survey to institute members and selected members of the project consortium. 
In this iteration, we got the feedback that answering the complete survey takes 30 minutes and thereby too much time. 
The reasons for that were too many questions for all the different working areas, the utilized kinds of models in these areas, and the tools associated with the models. To tackle this concern, the questions for which models are used in which work area and with which tool were changed into dependent questions. 
Consequently, a participant first chooses his/her work areas and afterward selects the kinds of models used in each selected work area. Likewise, we only ask participants for a selected kind of model to name the tools employed and rate their importance in the different phases of automotive development. 
In addition, in both preliminary studies, we addressed feedback regarding misleading formulations or spelling errors.
To avoid bias, participants in the two preliminary studies were excluded from the final questionnaire.

\subsection{Questionnaire Design}
The \dt in the SofDCar project includes designing a \dt information management layer that manages car configurations and information. 
To formalize this information and variants, models should be used. 
For this purpose, we designed a questionnaire to get an overview of employed models and associated tools in the consortium. 
Concerning their employment by \dts, we furthermore, wanted to gain the importance of purposes, properties, and development phases that \dts should fulfill or be used in. 
Therefore, we split the questionnaire into three parts comprising in total of 54 questions. 
Since the project consortium consists of experts from different domains, our first question asked for the scientific background of the participant, \ie cybernetics, business administration, physics, mechatronics, mechanical engineering, electrical engineering, or computer science. 
The first part of the questionnaire focuses on questions regarding the purposes, properties, and important development phases of \dts; the second part focuses on modeling techniques and tools; and \sosym{the third part is concerned with the use and reuse of informal models.}

\subsubsection{Properties, Purposes and Phases important for Digital Twins}
The first part of the survey started with questions concerning the \dt. 
We asked for the agreement on properties of a \dt using a four-point Likert scale ranging from disagree completely (-{}-), rather disagree (-) to rather agree (+) and completely agree (++). 
Participants could propose additional properties in an open text field. 
After the properties, we asked for the importance of purposes.
Here the participants, again, could rate on a four-point Likert scale whether they consider a purpose unimportant (-{}-), rather unimportant (-), rather important (+) or very important (++). 
The participants could enter additional purposes in a text field. 
The last question of the first part of the questionnaire was about the importance of the \dt in different development phases. 
Here the same four-point Likert scale as before for importance was used. 
\subsubsection{Models and Tools in the Automotive Industry}
For the second part of the questionnaire, we commenced by asking for the utilization of models in various work areas in automotive development.
Here, the participants could select multiple work areas using checkboxes, whereas a text field permitted participants to provide additional work areas. 
Depending on the selected work area, one additional question asked participants to name the kinds of models they employ in each work area. 
Again, checkboxes were used, and additional model kinds could be provided in a text field. 
Depending on the model kind a participant selected, two model-kind-specific questions were asked for each model kind.
First, which tools the participant uses to work on this model kind.
Second, how important they consider each model kind for the different phases of development, employing the same Likert scale as before ranging from unimportant to very important.
\sosym{
\subsubsection{Informal Models in the Automotive Industry}\label{sec:methodology-informal}
In the third part, we ask participants about the prevalence and utilization of informal models, e.g., hand-drawn on whiteboards or digitally sketched in PowerPoint\texttrademark, used in the automotive industry.
In detail, participants should rate how often informal models were used for the purpose of communication, documentation, design of the \dt, and requirements analysis of the \dt.
Moreover, they should rate how frequently informal models were utilized in each phase of development.
Furthermore, participants should state how often an informal model is stored as a digital sketch (e.g., in PowerPoint), as a photograph, as a transcribed model (manually transferred using a modeling tool) or in place (e.g., on the whiteboard).
For the above questions, we provided a five-point Likert scale ranging from very often (++), rather often (+), rather rarely (-) to rarely (-{}-) and never ($\emptyset$). Additionally, participants could enter custom purposes and development phases in a text field.
Next, participants should select all model kinds they sketch or state custom model kinds. To limit the number of selectable kinds, we picked those model kinds selected in the prestudies.
Finally, we tasked participants to rate their agreement to the following seven statements regarding issues using and persisting informal models:
\begin{enumerate*}[label=(\arabic*)]
    \item \textit{informal models are not reused because they are not stored;}
    \item \textit{informal models are not reused although they were stored;}
    \item \textit{photos of informal models were taken and reused later;}
    \item \textit{photos of informal models were taken yet never reused;}
    \item \textit{manually transcribing models was time consuming;}
    \item \textit{manually transcribing models were useful for later tasks;}
    \item \textit{stored informal models were difficult to find/retrieve.}
\end{enumerate*}
Again, we employ the same four-point Likert scale for agreement and provide an extra field for participants to list other encountered issues wrt.~re-/using informal models.
\revone{The statements arose from discussions within the project that date before the conduction of the study. We formulated these hypotheses to assess their validity.}
}

\subsection{Selection of Participants}
To reach as many participants from diverse work areas as possible, we invited the participants by email via an internal mailing list of 189 members of the SofDCar project. 
To improve the response rate, we implemented two deadline extensions from the initial one-month-long time frame with 2 weeks of additional response time. 
Together with the deadline extension, we sent a mail to inform and remind the members of the consortium of the survey. 


%
\begin{figure}
    \begin{subfigure}[b]{0.70\linewidth}
    \centering
    \begin{tikzpicture}[scale=.8]
\begin{axis}[ xbar, xmajorgrids, enlargelimits=0.1,
    width=.8\linewidth,height=6cm,
    nodes near coords, 
    xmin=0, xmax=23,
    ytick={0,1,...,10}, yticklabels={Sales/distribution,Production,Infotainment,Vehicle requirements,Chassis,Process development,Electronics,Electrics,Autonomous systems,Power train,Diagnostics (*)},
    every axis plot/.append style={bar shift=0pt}
    ]
  \addplot coordinates { 
    (1,0)
    (8,1)
    (5,2)
    (11,3)
    (1,4)
    (20,5)
    (11,6)
    (5,7)
    (10,8)
    (4,9)
  };
  \addplot[gray!25!black,fill=gray!50!white] coordinates {
    (2,10)
  };
\end{axis}
\end{tikzpicture}
    \vspace{.25cm}
    \subcaption{Participants{'} work areas}
    \label{fig:hist-work-areas}
    \end{subfigure}
    \begin{subfigure}[b]{0.28\linewidth}
    \centering
    \begin{tikzpicture}[scale=.8]
	\begin{axis}[width=1\linewidth, height=6cm, xmin=-0.5, xmax=0.5,
		   ymin=0, ymax=7,
		   ybar, ymajorgrids, enlargelimits=0.1,
		   legend style={anchor=north east,legend columns=6},
		   scaled ticks=false, tick label style={/pgf/number format/fixed},
		   xtick={0}, x tick style={draw=none},
          xticklabels={Participant},
          ytick={0,1,...,11},
          ylabel={\textbf{Number of work areas}}
		  ]

    \boxplot[blue]{0}{1}{1}{2}{1}{3}
    \addplot[only marks,mark=o,fill=none,blue,forget plot] coordinates {(0,4) (0,5) (0,7)};
    
	\end{axis}
\end{tikzpicture}    
    \subcaption{Per participant}
    \label{fig:boxplot-work-areas-per-participant}
    \end{subfigure}
    \caption{Number of participants in different work areas (a),  where the * indicates answers from the free-form field, and a box plot highlighting the distribution of work areas per participant (b).}
    \label{fig:work-areas}
\end{figure}
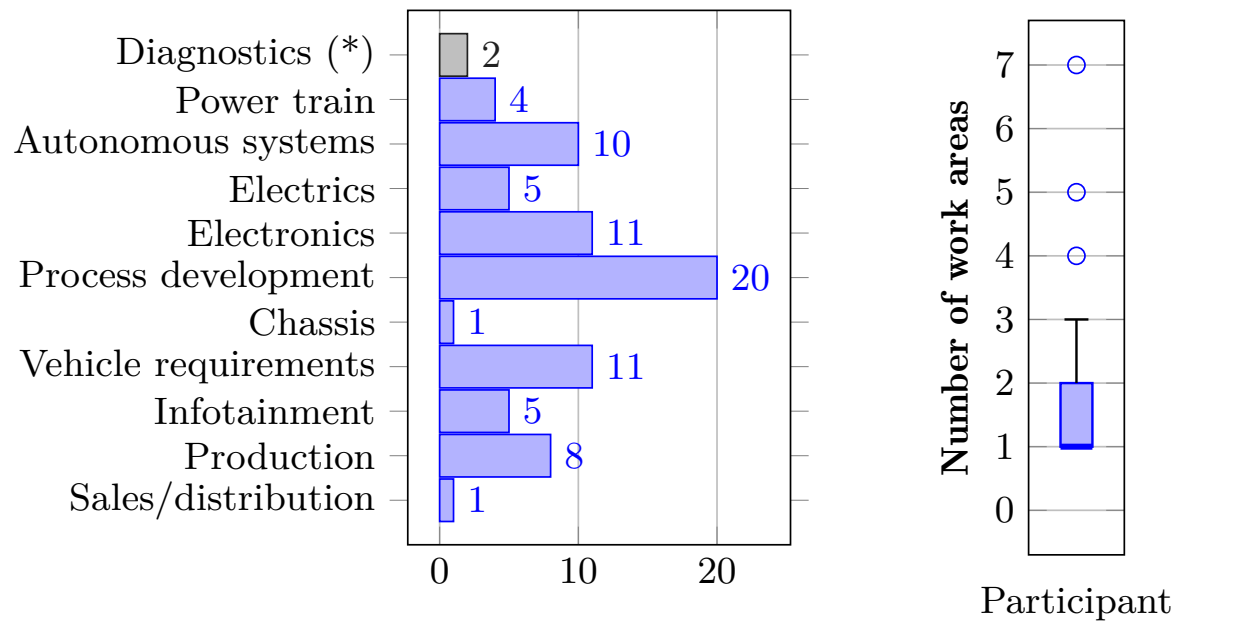

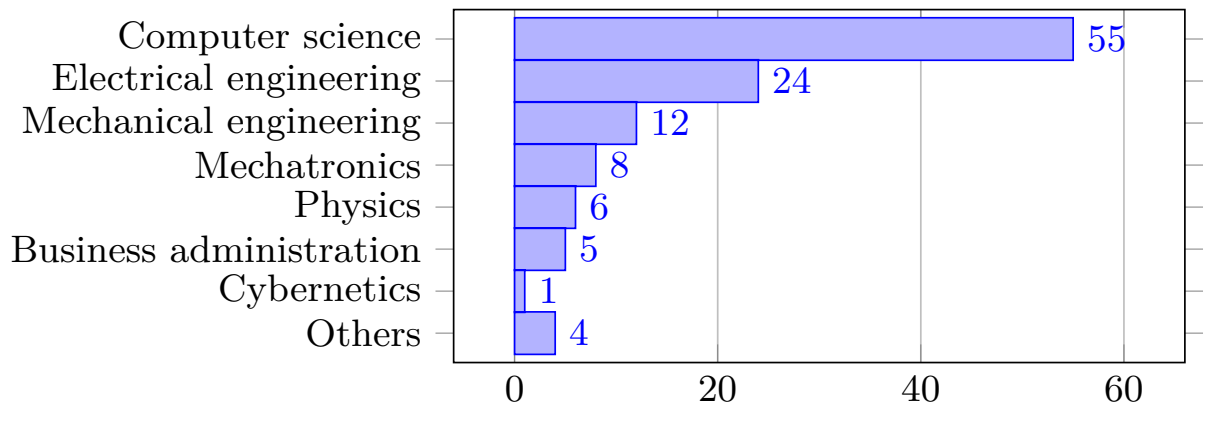
\begin{figure}
    \centering
    \begin{tikzpicture}[scale=.8]
\begin{axis}[ xbar, xmajorgrids, enlargelimits=0.1,
    width=.90\linewidth,height=4.5cm,
    nodes near coords, 
    xmin=0, xmax=60,
    ytick={0,...,7},
    yticklabels={Others, Cybernetics, Business administration, Physics, Mechatronics, Mechanical engineering, Electrical engineering, Computer science}
    ]
  \addplot coordinates { 
    (55,7)
    (24,6)
    (12,5)
    (8,4)
    (6,3)
    (5,2)
    (1,1)
    (4,0)
    };
\end{axis}
\end{tikzpicture}
    \caption{Overview of the professional backgrounds of the participants, whereas one participant could select multiple backgrounds.}
    \label{fig:participant-background}
\end{figure}

\section{Results}\label{sec:results}
This section presents the results of our survey.
In total, we had 96 participants, of which 43 completed the questionnaire.
\autoref{fig:participant-background} shows the scientific background of all participants of our survey.
For incomplete survey responses, we took only the available answers and did not include the missing answers in our data analysis.
The majority of participants were computer scientists (55), followed by electrical (25) and mechanical engineers (11).
The participants work in a diverse set of areas, shown in~\autoref{fig:hist-work-areas}, ranging from \textit{sales and distribution} and \textit{chassis} over \textit{power train} and \textit{infotainment} to \textit{vehicle requirements} and \textit{autonomous systems}.
Overall, most participants work on improving the automotive development processes and only work in up to two work areas (\cf \autoref{fig:boxplot-work-areas-per-participant}). Including automotive diagnostics, named as an additional work area, we reached participants of 11 distinct work areas.

\subsection{RQ1: How is the \dt understood in the automotive industry?}
\begin{figure*}
    \centering
    \begin{subfigure}{0.22\linewidth}
        \centering
        \begin{tikzpicture}[scale=0.8]
	\begin{axis}[width=1.2\linewidth, height=4cm, xmin=-0, xmax=4,
		   ymin=1, ymax=4,
		   ybar, ymajorgrids, enlargelimits=0.1,
		   legend style={anchor=north east,legend columns=6},
		   scaled ticks=false, tick label style={/pgf/number format/fixed},
		   xtick={0,1,...,4}, 
          xticklabels={Conception,Development,Production,Usage \& support,Retirement},
          x tick label style={rotate=90,anchor=east},
          ytick={1,2,3,4},
		   yticklabels={-~-,-,+,++},
		  ]

    \boxplot[blue]{0}{3}{2}{3}{1}{4}
    \boxplot[blue]{1}{4}{3}{4}{2}{4}    
    \boxplot[blue]{2}{3}{3}{4}{2}{4}
    \addplot[only marks,mark=o,fill=none,blue,forget plot] coordinates {(3,2) (3,3)};
    \boxplot[blue]{3}{4}{4}{4}{4}{4}
    \boxplot[blue]{4}{3}{2}{3}{1}{4}
    
	\end{axis}
\end{tikzpicture}
        \subcaption{Phases of development}\label{fig:rq1_a_boxplot}
    \end{subfigure}
    \begin{subfigure}{0.44\linewidth}
        \centering
        \begin{tikzpicture}[scale=0.8]
	\begin{axis}[width=1.2\linewidth, height=4cm, xmin=-0, xmax=11,
		   ymin=1, ymax=4,
		   ybar, ymajorgrids, enlargelimits=0.1,
		   legend style={anchor=north east,legend columns=6},
		   scaled ticks=false, tick label style={/pgf/number format/fixed},
		   xtick={0,1,...,11}, 
          xticklabels={Simulates,Virtual rep.,Set of models,Services,Contains data,Bi-directional,Collects data,Influences,Multiple DTs,As-designed,As-manufactured,As-operated},
          x tick label style={rotate=90,anchor=east},
          ytick={1,2,3,4},
		   yticklabels={-~-,-,+,++},
		  ]

    \boxplot[red]{0}{3}{3}{4}{2}{4}
    \boxplot[red]{1}{4}{3}{4}{3}{4}    
    \boxplot[red]{2}{3}{3}{4}{2}{4}
    \addplot[only marks,mark=o,fill=none,red,forget plot] coordinates {(3,1) (3,2) (3,4)};
    \boxplot[red]{3}{3}{3}{3}{3}{3}
    \boxplot[red]{4}{3}{3}{4}{2}{4}
    \boxplot[red]{5}{4}{3}{4}{2}{4}
    \boxplot[red]{6}{4}{3}{4}{2}{4}
    \boxplot[red]{7}{3}{2}{4}{1}{4}
    \boxplot[red]{8}{3}{2}{4}{1}{4}
    \boxplot[red]{9}{3}{2}{3}{1}{4}
    \boxplot[red]{10}{3}{2}{3}{1}{4}
    \boxplot[red]{11}{4}{3}{4}{2}{4}
    
	\end{axis}
\end{tikzpicture}
        \subcaption{Properties}\label{fig:rq1_b_boxplot}
    \end{subfigure}
    \begin{subfigure}{0.28\linewidth}
        \centering
        \begin{tikzpicture}[scale=0.8]
	\begin{axis}[width=1.2\linewidth, height=4cm, xmin=-0, xmax=6,
		   ymin=1, ymax=4,
		   ybar, ymajorgrids, enlargelimits=0.1,
		   legend style={anchor=north east,legend columns=6},
		   scaled ticks=false, tick label style={/pgf/number format/fixed},
		   xtick={0,1,...,6}, 
          xticklabels={Monitoring,Validation,Homologation,Coordinating,Prediction,Process opti.,Risk assessment},
          x tick label style={rotate=90,anchor=east},
          ytick={1,2,3,4},
		   yticklabels={-~-,-,+,++},
		  ]

    \boxplot[orange]{0}{4}{3}{4}{2}{4}
    \addplot[only marks,mark=o,fill=none,orange,forget plot] coordinates {(1,1)};
    \boxplot[orange]{1}{4}{3}{4}{2}{4}   
    \boxplot[orange]{2}{3}{2}{4}{1}{4}
    \boxplot[orange]{3}{3}{2}{3}{1}{4}
    \addplot[only marks,mark=o,fill=none,orange,forget plot] coordinates {(4,2)};
    \boxplot[orange]{4}{4}{3.5}{4}{3}{4}
    \addplot[only marks,mark=o,fill=none,orange,forget plot] coordinates {(5,1)};
    \boxplot[orange]{5}{3}{3}{4}{2}{4}
    \addplot[only marks,mark=o,fill=none,orange,forget plot] coordinates {(6,1)};
    \boxplot[orange]{6}{3}{3}{4}{2}{4}
    
	\end{axis}
\end{tikzpicture}
        \subcaption{Purposes}\label{fig:rq1_c_boxplot}
    \end{subfigure}
    \caption{Box plots of the perceived importance of \dts in the different phases of development (a), of the properties of \dts (b), and of purposes by employing \dts (c). The importance ranges from \textit{unimportant} ($-\,-$), via \textit{rather unimportant} ($-$) and \textit{rather important} (+) to \textit{very important} (++).}\label{fig:rq1_boxplot}
\end{figure*}
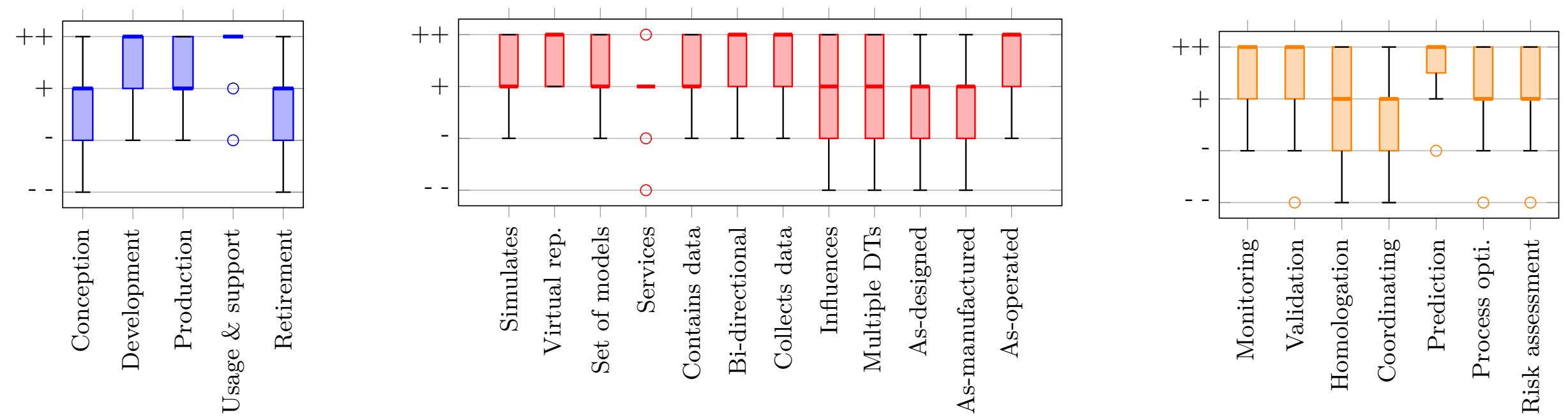

\subsubsection{RQ1.1: For which phases of automotive development are \dts important?}

To answer the question of how important the participants consider \dts in different phases of automotive development, we asked them to rate the relevance of \dts in the phases \textit{concept}, \textit{development}, \textit{production}, \textit{usage \& support}, and \textit{retirement}.
The answers ranged from \textit{unimportant} to \textit{very important}.
The box plot in \autoref{fig:rq1_a_boxplot} shows our results.
We can see that the \dt is considered important in all phases of automotive development.
Nevertheless, the results show that it is less important during the \textit{concept} and \textit{retirement} phase.
Furthermore, it is considered very important in the \textit{usage \& support} phase as well as in the \textit{development} phase.

\subsubsection{RQ1.2: What are desired properties of \dts?}
To answer the question of which properties a \dt should have, we asked the participants to rate the importance of various properties via a Likert scale from unimportant to very important. We selected the following properties based on current literature and discussions with our partners in the project:
\begin{inparaenum}[(1)]
    \item \textit{Simulates}: the \dt simulates the actual \cps.
    \item \textit{Virtual representation}: the \dt is a virtual representation of the actual \cps.
    \item \textit{Set of models}: the \dt encapsulates a set of models.
    \item Services: the \dt provides services for the \cps.
    \item \textit{Contains data}: the \dt contains data that is relevant for the \cps.
    \item \textit{Bidirectional}: a bi-directional connection between the \dt and the \cps.
    \item \textit{Collects data}: the \dt can collect current data of the \cps.
    \item \textit{Influences}: the ability of the \dt to influence the \cps.
    \item multiple \dts: one \cps can have multiple \dts.
    \item \textit{As-designed}: the \dt describes the \cps in its state at design time.
    \item \textit{As-manufactured}: the \dt describes how the \cps is manufactured.
    \item \textit{As-operated}: the \dt describes how the \cps behaves during runtime.
\end{inparaenum}
The box plot in \autoref{fig:rq1_b_boxplot} shows the results regarding the different possible properties.
We can see that the participants consider all properties as rather important or very important (median).
The results show that the properties of \textit{virtual representation}, \textit{as-operated}, \textit{collects data}, and \textit{bi-directional synchronization} are considered very important.

\subsubsection{RQ1.3: What are desired purposes of using \dts?}

\dts are employed for achieving a multitude of purposes.
Based on the existing literature and discussion with industry partners in our consortium, we provided the following purposes in our questionnaire: \begin{inparaenum}[(1)]
    \item \textit{Monitoring}: the analysis of runtime and historical data emitted from the \cps.
    \item \textit{Validation}: the \dt can test the qualities and functionalities of the \cps.
    \item \textit{Homologation}: helps car manufacturers to get admission by authorities for their car, its components and software.
    \item \textit{Coordinating}: the \dt can coordinate and control multiple \cpss, \eg for fleet management.
    \item \textit{Prediction}: the \dt can make predictions on the \cps, \eg for erroneous behavior or the next maintenance required.
    \item \textit{Process optimization}: the \dt can improve development, sales and operational processes.
    \item \textit{Risk assessment}: the \dt can point out possible risks depending on the phases of development in automotive.
\end{inparaenum}
The participants could rate how important they consider the respective purposes based on a Likert scale ranging from unimportant to very important.
(1) For \textit{monitoring}, the answers ranged from 2 rather not important to 4 very important.
However, the median and the majority of participants considered that this is a very important purpose of a \dt.
(2) For the purpose of \textit{validation}, the median and majority considered it very important, too.
Only a few participants voted that this \textit{validation} is not an important purpose of a \dt.
(3) For the purpose of \textit{homologation} the answers ranged from not important to very important. The median, however, is rather important.
(4) This is true for the purpose of \textit{coordination}, too.
(5) \textit{Prediction} has the most answers considering it as very important and no answer considering it not important at all.
(6) For \textit{process optimization}, the median of answers considers this purpose as rather important. While most answers were between rather important and very important, some participants replied with not important and rather not important.
(7) For the purpose of risk assessment,  the results are the same as for \textit{process optimization}.

\subsubsection{RQ1.4: How do these purposes change in relation to different phases of automotive development?}\label{sec:results-rq1.4}
\begin{figure}[tb]
    \centering
    \includegraphics[width=\linewidth]{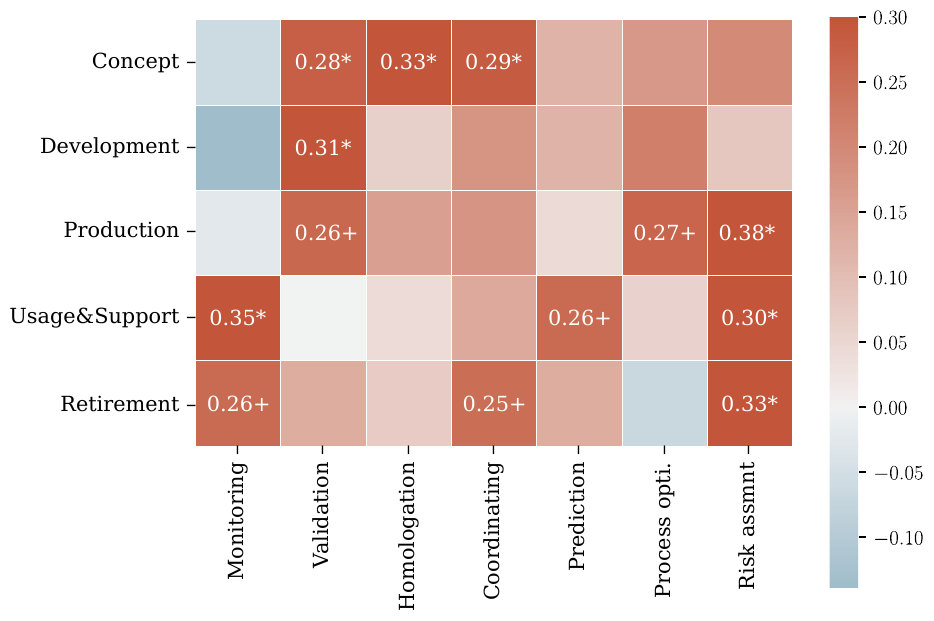}
    \caption{\revone{Heatmap of the correlations between phases of development and purposes for \dts. Cells show the Pearson correlation, whereas the hue indicates a positive or negative correlation. Strong significance ($p<0.05$) is indicated with a \texttt{*} and weak significance ($p<0.10$) with a \texttt{+}.}}
    \label{fig:rq1_d_heatmap}
\end{figure}
To answer the question of which \dt purposes are relevant in specific development phases, we related the answers to both questions to one another, resulting in the heatmap depicted in \autoref{fig:rq1_d_heatmap}.
It shows that the purpose of \textit{monitoring} is especially relevant in the post-production phases of \textit{support}~\revone{($p=.011$)} and \textit{retirement}~\revone{($p=.066$)}.
In contrast, \textit{validation} is important in the pre-operational phases of \textit{conception}~\revone{($p=.048$)}, \textit{development}~\revone{($p=.026$)} and \textit{production}~\revone{($p=.064$)}.
\textit{Homologation} is at all relevant in the \textit{conception} phase~\revone{($p=.023$)}.
\textit{Coordinating} as a purpose of the \dt is considered relevant in all development phases but especially in the \textit{conception}~\revone{($p=.047$)} and \textit{retirement}~\revone{($p=.081$)} phases.
The purpose \textit{prediction} is a purpose that is mostly relevant in the \textit{usage \& support} phase~\revone{($p=.068$)}.
\textit{Process optimization} is highly relevant to the production process and, thus, there is a strong relation between this purpose and the \textit{production} phase~\revone{($p=.06$)}.
\textit{Risk assessment} has a strong correlation in the phases of \textit{production}~\revone{($p=.006$)}, \textit{support}~\revone{($p=.03$)}, and \textit{retirement}~\revone{($p=.018$)}.



\subsection{RQ2: Which modeling languages and modeling tools are currently employed in the automotive industry?}

\subsubsection{RQ2.1: Which kinds of models are important during automotive development?}
\begin{figure}
    \begin{subfigure}[b]{0.60\linewidth}
    \centering
    \begin{tikzpicture}[scale=.8]
\begin{axis}[ xbar, xmajorgrids, enlargelimits=0.1,
    width=.8\linewidth,height=6.5cm,
    nodes near coords, 
    xmin=0, xmax=30,
    ytick={0,...,11},
    yticklabels={Circuit models,CAD models,E/E architecture,Source code,Data schema,Math/phy. models,FE models,Simulation models,Business proc.~models,Development proc.~models,Variability models,Deployment models (*)},
    every axis plot/.append style={bar shift=0pt}
    ]
  \addplot coordinates { 
    (2,0)
    (5,1)
    (14,2)
    (25,3)
    (16,4)
    (15,5)
    (5,6)
    (21,7)
    (19,8)
    (10,9)
    (13,10)
  };
  \addplot[gray!25!black,fill=gray!50!white] coordinates {
    (2,11)
  };
\end{axis}
\end{tikzpicture}
    \vspace{-.25cm}
    \subcaption{Number of different model kinds}
    \label{fig:hist-model-kind}
    \end{subfigure}
    \begin{subfigure}[b]{0.28\linewidth}
    \centering
    \begin{tikzpicture}[scale=.8]
	\begin{axis}[width=1\linewidth, height=6.5cm, xmin=-0.5, xmax=0.5,
		   ymin=0, ymax=11,
		   ybar, ymajorgrids, enlargelimits=0.1,
		   legend style={anchor=north east,legend columns=6},
		   scaled ticks=false, tick label style={/pgf/number format/fixed},
		   xtick={0}, x tick style={draw=none},
          xticklabels={Participant},
          ytick={0,1,...,11},
          ylabel={\textbf{Number of model kinds}}
		  ]

    \boxplot[blue]{0}{3}{1.75}{4}{1}{6}
    \addplot[only marks,mark=o,fill=none,blue,forget plot] coordinates {(0,8) (0,9) (0,11)};
    
	\end{axis}
\end{tikzpicture}
    \subcaption{Per participant}
    \label{fig:boxplot-model-kind-per-participant}
    \end{subfigure}
    \caption{Number of different model kinds employed by participants (a), where the * indicates answers from the free-form field, and a box plot highlighting the distribution of model kinds used per participant (b).}
    \label{fig:hist-model-kinds}
\end{figure}
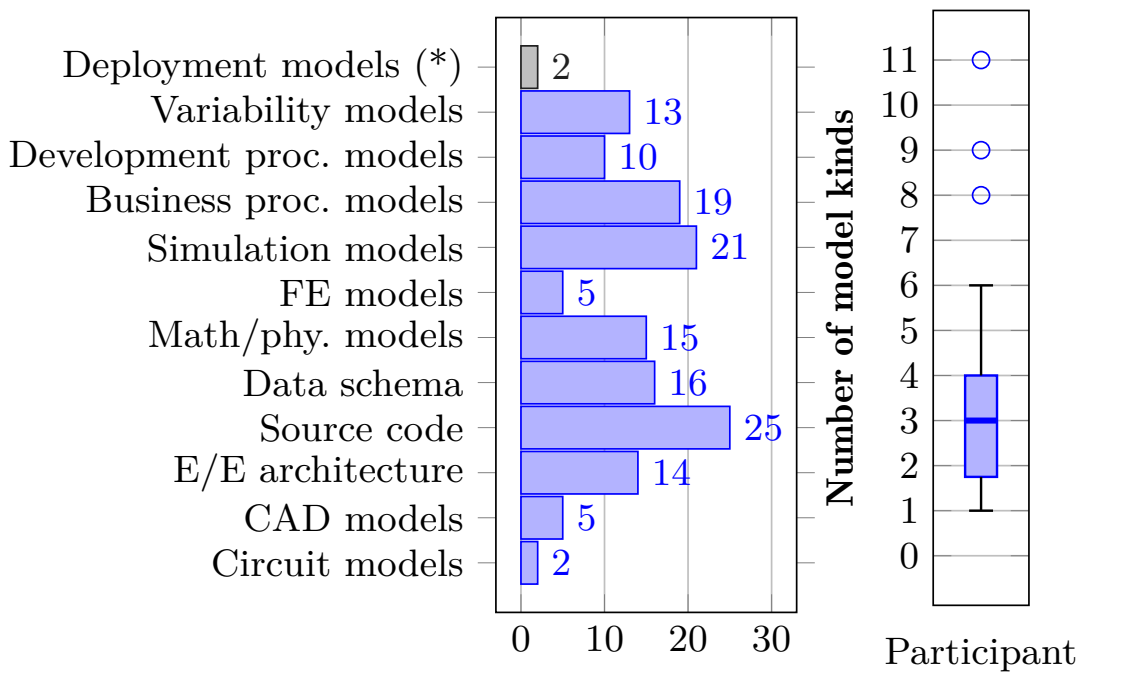

As we were also interested in the kinds of models and modeling tools currently employed for automotive development, we first asked each participant to pick their work areas.
From the provided work areas, participants worked in 10 different areas, i.e., power train, autonomous systems, electrics, electronics, process development, chassis, vehicle requirements, infotainment, production, sales/distribution.
Overall participants, all 12 model kinds were stated to be employed within automotive development, as shown in~\autoref{fig:hist-model-kind}:
\begin{inparaenum}[(1)]
    \item \textit{Variability models}: represent all possible variants of a system, usually by means of features and constraints.
    \item Development process models: model all aspects of a development process, including project schedule and organization, employed to supervise the development process.
    \item \textit{Business process models}: represent business processes by means of activities, events, and flows; permits their analysis, management, and automation.
    \item \textit{Simulation models}: specify complex systems with continuous and discrete elements, such that the system's behavior can be simulated.
    \item \textit{Finite element (FE) models}: represent the physical shape and properties of objects to apply the finite element method~\cite{huebner2001finite} used for structural analysis, heat transfer or body deformation.
    \item \textit{Mathematical/physical models}: capture physical phenomena and dynamic behavior of a \cps by means of mathematical equations.
    \item \textit{Data schema}: specify the layout and type of data used, stored, and/or shared within a system.
    \item \textit{Source code}: is a textual, human-readable representation of a machine-executable program.
    \item \textit{Electrical/electronic (E/E) architecture}: models the electrical and electronic hardware, the software components, and the wiring of a \cps.
    \item \textit{Computer-Aided Design (CAD) models}: capture the physical shape and composition of 3D objects.
    \item \textit{Circuit models}: describe an electronic circuit used to design and emulate electronic hardware.
\end{inparaenum}
In addition to the 11 model kinds we provided, two participants declared \textit{deployment models} as an additional model kind that denotes the run-time configuration of software components running on processing nodes used to deploy or reconfigure a \cps.
%
The most common model kinds were \textit{source code} (25) followed by \textit{simulation models} (21) and \textit{business process models} (19).
In contrast, \textit{FE models} (5) and \textit{circuit models} (2) were the least common model kinds employed.
Moreover, we determined how many different model kinds each participant named. The result is depicted in~\autoref{fig:boxplot-model-kind-per-participant}, whereas the median number of different model kinds per participant is three. Notably, one participant selected all model kinds, which is a clear outlier.

\begin{figure}
    \centering
    \input{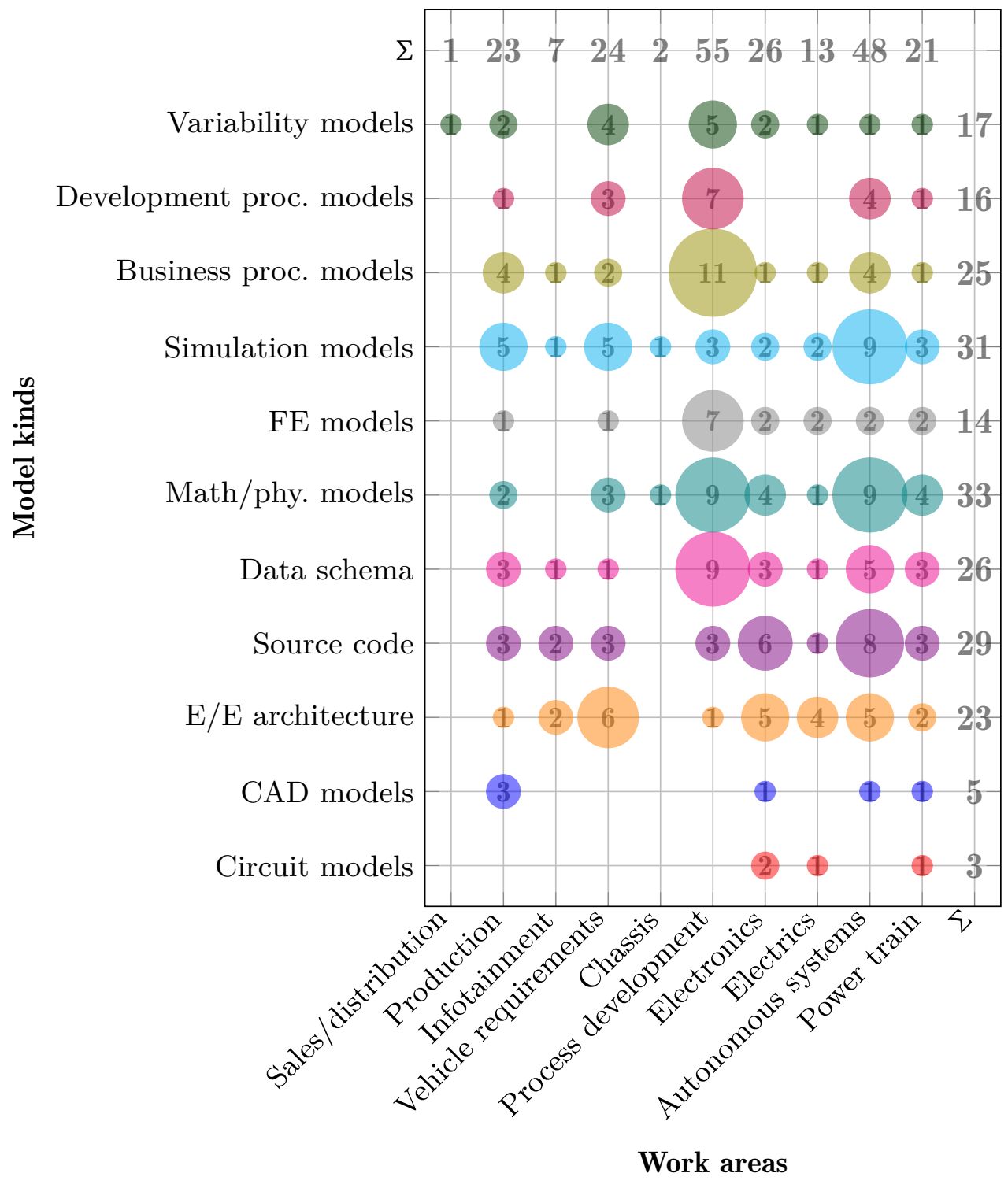}
    \caption{Bubble chart relating the model kind (y-axis) to the various work areas of automotive development (x-axis).}
    \label{fig:bubble-aspect-models}
\end{figure}

To further distinguish, in which work area which model kind was used, the bubble chart in \autoref{fig:bubble-aspect-models} indicates, which model kinds are employed in which work area.
Please note that several participants stated to work in multiple working areas, and, in turn, named the same model kind in different work areas. In the bubble chart, these mentions are counted separately.
The data indicates that, except for \textit{sales \& distribution}, each work area employs a wide variety of different model kinds. Surprisingly, we found that the work area of \textit{process development} also employs \textit{FE models} and \textit{mathematical/physical~models}.

\subsubsection{RQ2.2: How important are which models in the phases of automotive development}
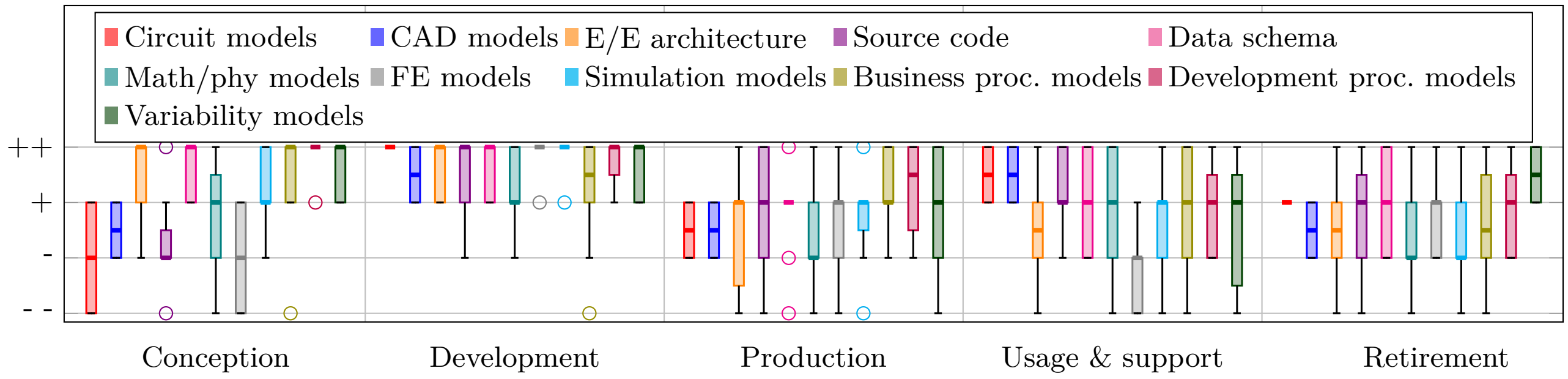
\begin{figure*}
    \centering
    \begin{tikzpicture}[scale=1]
	\begin{axis}[width=\linewidth, height=5cm, xmin=-0.5, xmax=58.5,
		   ymin=0.9, ymax=6.5,
		   ybar, ymajorgrids, enlargelimits=0.01,
		   legend style={anchor=north east,legend columns=5},
		   scaled ticks=false, tick label style={/pgf/number format/fixed},
		   xtick={0,1,...,58}, x tick style={draw=none},
          xticklabels={,,,,,Conception,,,,,,,,,,,,Development,,,,,,,,,,,,Production,,,,,,,,,,,,Usage \& support,,,,,,,,,,,,,Retirement},
          x tick label style={rotate=0,anchor=north},
          xminorgrids, minor xtick={11,23,35,47},
          ytick={1,2,3,4},
		   yticklabels={-~-,-,+,++},
          legend cell align={left}
		  ]

    \boxplot[circuit]{0}{2}{1}{3}{1}{3}
    \boxplot[cad]{1}{2.5}{2}{3}{2}{3}
    \boxplot[eea]{2}{4}{3}{4}{2}{4}    
    \addplot[only marks,mark=o,fill=none,code,forget plot] coordinates {(3,1) (3,4)};
    \boxplot[code]{3}{2}{2}{2.5}{2}{3}    
    \boxplot[data]{4}{4}{3}{4}{3}{4}
    \boxplot[math]{5}{3}{2}{3.5}{1}{4}
    \boxplot[fem]{6}{2}{1}{3}{1}{3}
    \boxplot[simulation]{7}{3}{3}{4}{2}{4}
    \addplot[only marks,mark=o,fill=none,workflow,forget plot] coordinates {(8,1)};    
    \boxplot[workflow]{8}{4}{3}{4}{3}{4} 
    \addplot[only marks,mark=o,fill=none,process,forget plot] coordinates {(9,3)};    
    \boxplot[process]{9}{4}{4}{4}{4}{4}    
    \boxplot[variability]{10}{4}{3}{4}{3}{4}

    \boxplot[circuit]{12}{4}{4}{4}{4}{4}
    \boxplot[cad]{13}{3.5}{3}{4}{3}{4}
    \boxplot[eea]{14}{4}{3}{4}{3}{4}    
    \boxplot[code]{15}{4}{3}{4}{2}{4}
    \boxplot[data]{16}{4}{3}{4}{3}{4}
    \boxplot[math]{17}{3}{3}{4}{2}{4}
    \addplot[only marks,mark=o,fill=none,fem,forget plot] coordinates {(18,3)};     
    \boxplot[fem]{18}{4}{4}{4}{4}{4}
    \addplot[only marks,mark=o,fill=none,simulation,forget plot] coordinates {(19,3)};     
    \boxplot[simulation]{19}{4}{4}{4}{4}{4}  
    \addplot[only marks,mark=o,fill=none,workflow,forget plot] coordinates {(20,1)};
    \boxplot[workflow]{20}{3.5}{3}{4}{2}{4} 
    \boxplot[process]{21}{4}{3.5}{4}{3}{4}
    \boxplot[variability]{22}{4}{3}{4}{3}{4}

    \boxplot[circuit]{24}{2.5}{2}{3}{2}{3}
    \boxplot[cad]{25}{2.5}{2}{3}{2}{3}
    \boxplot[eea]{26}{3}{1.5}{3}{1}{4}    
    \boxplot[code]{27}{3}{2}{4}{1}{4}
    \addplot[only marks,mark=o,fill=none,data,forget plot] coordinates {(28,1) (28,2) (28,4)};    
    \boxplot[data]{28}{3}{3}{3}{3}{3}
    \boxplot[math]{29}{2}{2}{3}{1}{4}
    \boxplot[fem]{30}{3}{2}{3}{1}{4}
    \addplot[only marks,mark=o,fill=none,simulation,forget plot] coordinates {(31,1) (31,4)};     
    \boxplot[simulation]{31}{3}{2.5}{3}{2}{3}    
    \boxplot[workflow]{32}{3}{3}{4}{2}{4} 
    \boxplot[process]{33}{3.5}{2.5}{4}{2}{4}
    \boxplot[variability]{34}{3}{2}{4}{1}{4} 

    \boxplot[circuit]{36}{3.5}{3}{4}{3}{4}
    \boxplot[cad]{37}{3.5}{3}{4}{3}{4}
    \boxplot[eea]{38}{2.5}{2}{3}{1}{4}    
    \boxplot[code]{39}{3}{3}{4}{2}{4}
    \boxplot[data]{40}{3}{2}{4}{2}{4}
    \boxplot[math]{41}{3}{2}{4}{1}{4}
    \boxplot[fem]{42}{2}{1}{2}{1}{3}
    \boxplot[simulation]{43}{3}{2}{3}{1}{4}    
    \boxplot[workflow]{44}{3}{2}{4}{1}{4} 
    \boxplot[process]{45}{3}{2}{3.5}{2}{4}
    \boxplot[variability]{46}{3}{1.5}{3.5}{1}{4} 

    \boxplot[circuit]{48}{3}{3}{3}{3}{3}
    \boxplot[cad]{49}{2.5}{2}{3}{2}{3}
    \boxplot[eea]{50}{2.5}{2}{3}{1}{4}
    \boxplot[code]{51}{3}{2}{3.5}{1}{4}
    \boxplot[data]{52}{3}{2}{4}{2}{4}
    \boxplot[math]{53}{2}{2}{3}{1}{4}
    \boxplot[fem]{54}{3}{2}{3}{2}{4}
    \boxplot[simulation]{55}{2}{2}{3}{1}{4}
    \boxplot[workflow]{56}{2.5}{2}{3.5}{1}{4} 
    \boxplot[process]{57}{3}{2}{3.5}{2}{4}
    \boxplot[variability]{58}{3.5}{3}{4}{3}{4} 
    
    
    \addlegendimage{legend image with color=circuit!60};
    \addlegendentry{Circuit models};
    \addlegendimage{legend image with color=cad!60};
    \addlegendentry{CAD models};
    \addlegendimage{legend image with color=eea!60};
    \addlegendentry{E/E architecture};
    \addlegendimage{legend image with color=code!60};
    \addlegendentry{Source code};
    \addlegendimage{legend image with color=data!60};
    \addlegendentry{Data schema};
    \addlegendimage{legend image with color=math!60};
    \addlegendentry{Math/phy models};      
    \addlegendimage{legend image with color=fem!60};
    \addlegendentry{FE models};
    \addlegendimage{legend image with color=simulation!60};
    \addlegendentry{Simulation models};
    \addlegendimage{legend image with color=workflow!60};
    \addlegendentry{Business proc.~models};  
    \addlegendimage{legend image with color=process!60};
    \addlegendentry{Development proc.~models}; 
    \addlegendimage{legend image with color=variability!60};
    \addlegendentry{Variability models};      
	\end{axis}
\end{tikzpicture}
    \caption{Box plot of the importance of each model kind for the five phases of automotive development. The importance ranges from \textit{unimportant} (-{}-), via \textit{rather unimportant} (-) and \textit{rather important} (+) to \textit{very important} (+{}+). The number of answers per model kind differs, as participants could select to employ multiple model kinds.}
    \label{fig:boxplot-model-importance}
\end{figure*}

\begin{figure*}
    \centering
    \begin{subfigure}{0.555\linewidth}
    \includegraphics[width=\linewidth]{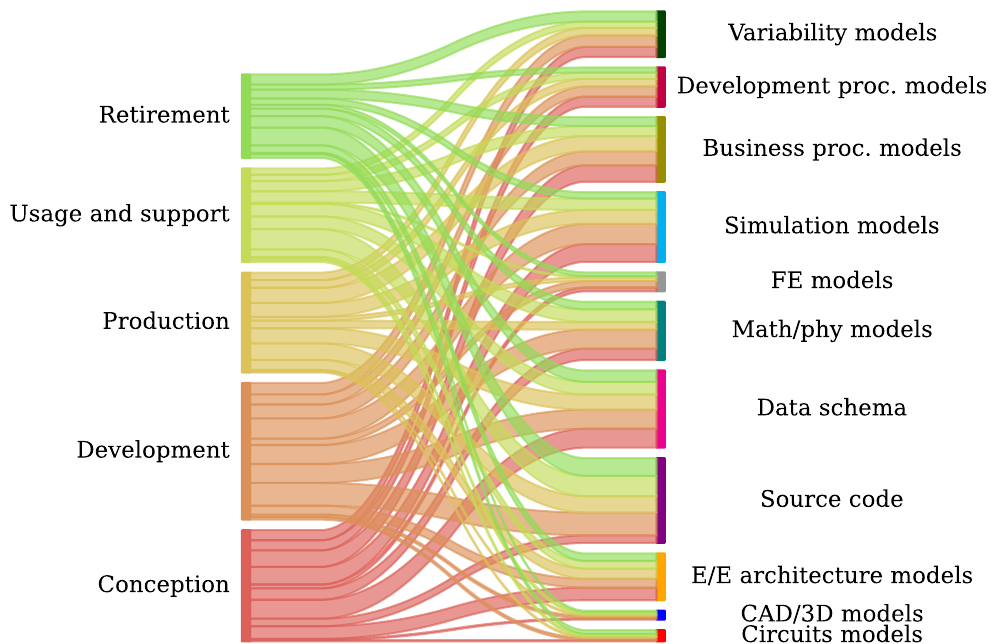}
    \subcaption{Model kinds employed in each phase}
    \label{fig:sankey-model-phase}
    \end{subfigure}
    \hspace{-0.9em}
    \begin{subfigure}{0.425\linewidth}
    \includegraphics[width=\linewidth]{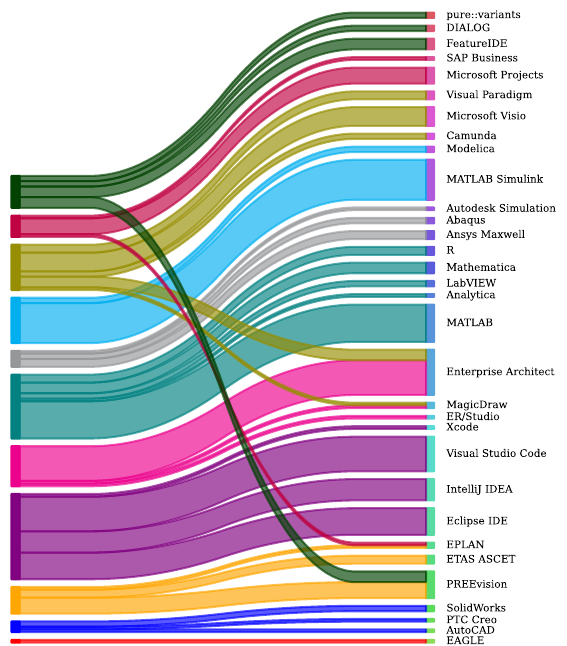}
    \subcaption{Tools employed to design model kind}
    \label{fig:sankey-model-tools}
    \end{subfigure}
    \caption{Sankey diagram relating the development phases to the various model kinds employed (a) as well as the tools used to model them. Thickness corresponds to the number of answers with \textit{rather important} or \textit{very important}. Participants named all model kinds they employ.}
    \label{fig:sankey-models}
\end{figure*}

To answer how relevant each model kind for each phase of automotive development is, each participant rated the importance of each selected model kind for each of the five phases.
The results, summarized in~\autoref{fig:boxplot-model-importance}, show that all model kinds are considered \textit{very} or \textit{rather important} in more than one phase.
For the \textit{conception phase}, \textit{E/E architecture models}, \textit{data schemas} and \textit{variability models} are considered \textit{very important} by most participants.
For the development phase, all model kinds are considered \textit{rather important} or \textit{very important} by most participants.
For the \textit{production} phase, only \textit{source code}, \textit{business process models}, \textit{development process models} and \textit{variability models} are considered as \textit{rather important} by most participants.
For the \textit{usage\& support phase}, in addition to the model kinds of the \textit{production} phase, \textit{circuit models}, \textit{CAD models}, \textit{data schemas}, and \textit{math./phy.~models} are considered \textit{rather important} by most participants.
Finally, for the \textit{retirement} phase, only \textit{source code}, \textit{data schemas}, \textit{development process models}, and \textit{variability models} are considered \textit{rather important} by most participants.
In essence, the participant answers indicate that \textit{data schemas}, \textit{business process models}, \textit{development process models}, and \textit{variability models} are considered \textit{very important} or \textit{rather important} throughout all phases of development when considering the median importance.
In contrast, the median importance of \textit{FE models} and \textit{math./phy.~models} varies per phase.
To illustrate the variate of model kinds considered important in each phase, \autoref{fig:sankey-model-phase} relates all development phases to the model kinds whereas the width of the line corresponds to the number participants considering the model kind as either \textit{rather} or \textit{very important}.
In sum, all model kinds are considered important in each of the five development phases.
This is true for \textit{CAD models} and \textit{FE models} despite the limited number of participants using them.

\subsubsection{RQ2.3: Which tools are used to create and maintain these models?}
In addition to the importance of model kinds, participants were asked to name the corresponding modeling tools or development environments.
In total, we collected individual 152 mentions of tools by participants resulting in 32 distinct modeling tools and development environments. Please note, that many participants stated multiple tools for each model kind.
The most common tools were \textsf{MatLab Simulink} for simulation models (15), \textsf{Enterprise Architect} for data schemas (14), \textsf{Visual Studio Code} for source code (14), and \textsf{MatLab} for mathematical/physical (math./phy.) models (14).
To highlight the tools used in the automotive domain, \autoref{fig:sankey-model-tools} shows the relation between model kinds and stated tools. Again, the width of the lines denotes the number of mentions.
Notably, though, a plethora of tools are currently in use whereas many only focus on a specific model kind.
Only four tools support multiple kinds of models, i.e., \textsf{MagicDraw}, \textsf{Enterprise Architect}, \textsf{PREEvision} and \textsf{EPLAN}.




\sosym{
\subsection{RQ3: How relevant are informal models in the automotive industry?}


\begin{figure*}
    \centering
    \begin{subfigure}{0.19\linewidth}
        \centering
        \begin{tikzpicture}[scale=0.8]
	\begin{axis}[width=1.2\linewidth, height=5cm, xmin=-0, xmax=3,
		   ymin=1, ymax=5,
		   ybar, ymajorgrids, enlargelimits=0.1,
		   legend style={anchor=north east,legend columns=6},
		   scaled ticks=false, tick label style={/pgf/number format/fixed},
		   xtick={0,1,2,3}, 
          xticklabels={Communication, Documentation, Design of DTs, \parbox{7.5em}{\raggedleft Requirements Analysis of DTs}},
          x tick label style={rotate=90,anchor=east},
          ytick={1,2,3,4,5},
		   yticklabels={$\emptyset$,-~-,-,+,++},
		  ]

    \boxplot[orange]{0}{4}{4}{5}{3}{5}
    \addplot[only marks,mark=o,fill=none,orange,forget plot] coordinates {(0,2)};

    \boxplot[orange]{1}{4}{4}{5}{3}{5}
    \addplot[only marks,mark=o,fill=none,orange,forget plot] coordinates {(1,1) (1,2)};

    \boxplot[orange]{2}{4}{3}{5}{2}{5}

    \boxplot[orange]{3}{4}{3}{5}{1}{5}

	\end{axis}
\end{tikzpicture}
        \subcaption{Purpose}\label{fig:rq3_a_boxplot}
    \end{subfigure}
    \begin{subfigure}{0.22\linewidth}
        \centering
        \begin{tikzpicture}[scale=0.8]
	\begin{axis}[width=1.2\linewidth, height=5cm, xmin=-0, xmax=4,
		   ymin=1, ymax=5,
		   ybar, ymajorgrids, enlargelimits=0.1,
		   legend style={anchor=north east,legend columns=6},
		   scaled ticks=false, tick label style={/pgf/number format/fixed},
		   xtick={0,1,...,4}, 
          xticklabels={Conception, Development, Production, \parbox{7.5em}{Usage \& support}, Retirement},
          x tick label style={rotate=90,anchor=east},
          ytick={1,2,3,4,5},
		   yticklabels={$\emptyset$,-~-,-,+,++},
		  ]

    \addplot[only marks,mark=o,fill=none,blue,forget plot] coordinates {(0,2)};
    \boxplot[blue]{0}{5}{4}{5}{4}{5}    
    \boxplot[blue]{1}{4}{4}{5}{3}{5}
    \boxplot[blue]{2}{3}{2}{4}{1}{5}
    \boxplot[blue]{3}{3}{2}{4}{1}{5}
    \boxplot[blue]{4}{3}{2}{4}{1}{5}
    
	\end{axis}
\end{tikzpicture}
        \subcaption{Phases of development}\label{fig:rq3_b_boxplot}
    \end{subfigure}
    \begin{subfigure}{0.19\linewidth}
        \centering
        \begin{tikzpicture}[scale=0.8]
	\begin{axis}[width=1.2\linewidth, height=5cm, xmin=-0, xmax=3,
		   ymin=1, ymax=5,
		   ybar, ymajorgrids, enlargelimits=0.1,
		   legend style={anchor=north east,legend columns=6},
		   scaled ticks=false, tick label style={/pgf/number format/fixed},
		   xtick={0,1,2,3}, 
          xticklabels={\parbox{7.5em}{Created digitally}, Transcribed, Photographed,\parbox{7.5em}{\raggedleft Retained on Whiteboard}},
          x tick label style={rotate=90,anchor=east},
          ytick={1,2,3,4,5},
		   yticklabels={$\emptyset$,-~-,-,+,++},
		  ]

    \boxplot[teal]{0}{4}{4}{5}{3}{5}
    \addplot[only marks,mark=o,fill=none,teal,forget plot] coordinates {(0,1) (0,2)};
    \boxplot[teal]{1}{3}{2}{4}{1}{5}   
    \boxplot[teal]{2}{4}{3}{5}{1}{5}
    \boxplot[teal]{3}{3}{2}{4}{1}{5}
    
	\end{axis}
\end{tikzpicture}
        \subcaption{Way of persisting}\label{fig:rq3_c_boxplot}
    \end{subfigure}
    \caption{\sosym{Box plots of the purposes informal models (a) serve, of their use in different phases of development (b), of the way they are persisted (c). The frequency ranges from \textit{never} ($\emptyset$), via \textit{rarely} (-{}-), \textit{rather rarely} (-), and \textit{rather often} (+) to \textit{very often}~(+{}+).}}\label{fig:rq3_boxplot}
\end{figure*}
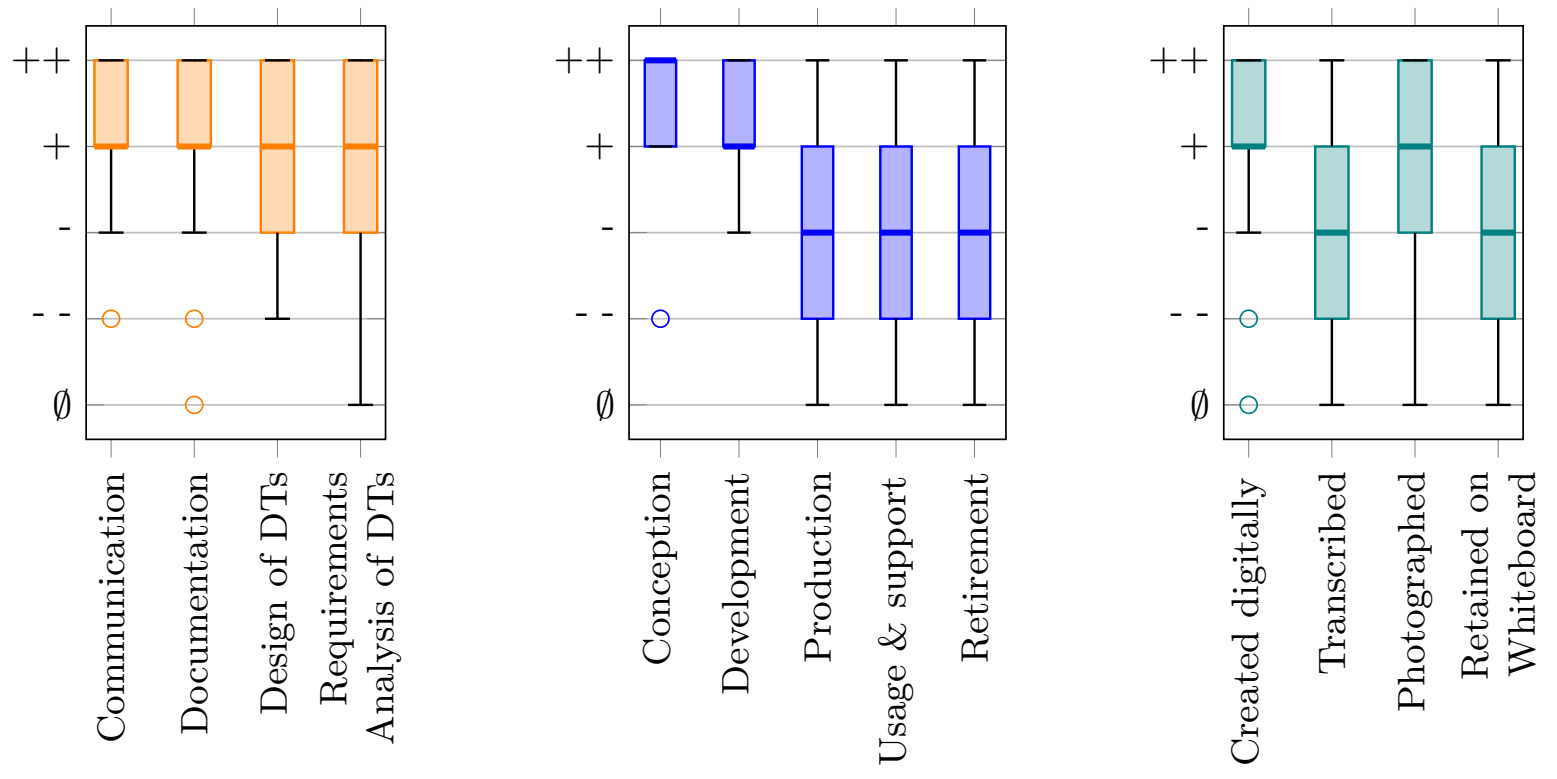

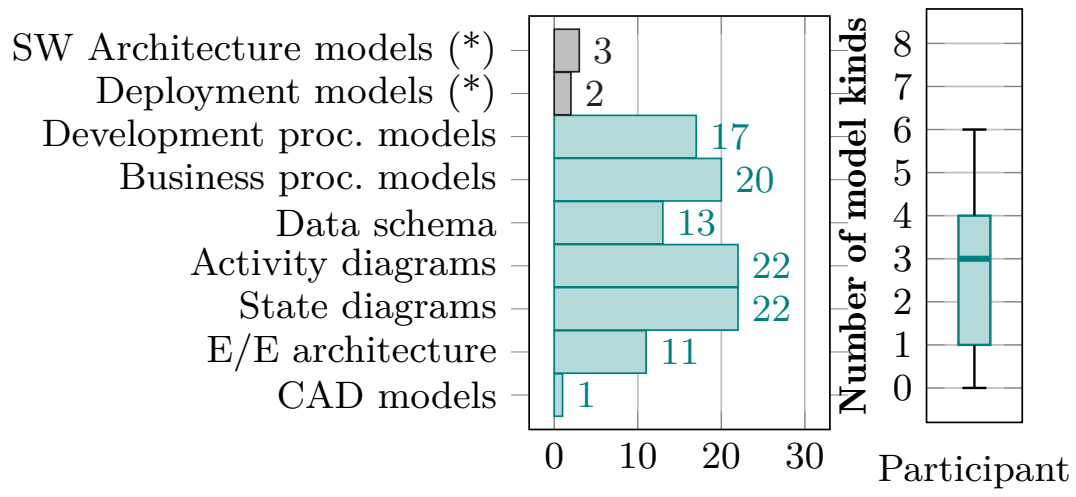
\begin{figure}
    \begin{subfigure}[b]{0.60\linewidth}
    \centering
    \begin{tikzpicture}[scale=.8]
\begin{axis}[ xbar, xmajorgrids, enlargelimits=0.1,
    width=.8\linewidth,height=5cm,
    nodes near coords, 
    xmin=0, xmax=30,
    ytick={0,...,8},
    yticklabels={CAD models, E/E architecture, State diagrams, Activity diagrams, Data schema, Business proc.~models, Development proc.~models, Deployment models (*),SW Architecture models (*)},
    every axis plot/.append style={bar shift=0pt}
    ]
  \addplot[teal,fill=teal!30!white] coordinates {
    (1,0)
    (11,1)
    (22,2)
    (22,3)
    (13,4)
    (20,5)
    (17,6)
  };
  \addplot[gray!25!black,fill=gray!50!white] coordinates {
    (2,7)
    (3,8)
  };
\end{axis}
\end{tikzpicture}
    \vspace{-.25cm}
    \subcaption{Number of different model kinds}
    \label{fig:hist-informal-model-kind}
    \end{subfigure}
    \begin{subfigure}[b]{0.28\linewidth}
    \centering
    \begin{tikzpicture}[scale=.8]
	\begin{axis}[width=1\linewidth, height=5cm, xmin=-0.5, xmax=0.5,
		   ymin=0, ymax=8,
		   ybar, ymajorgrids, enlargelimits=0.1,
		   legend style={anchor=north east,legend columns=6},
		   scaled ticks=false, tick label style={/pgf/number format/fixed},
		   xtick={0}, x tick style={draw=none},
          xticklabels={Participant},
          ytick={0,1,...,8},
          ylabel={\textbf{Number of model kinds}}
		  ]

    \boxplot[teal]{0}{3}{1}{4}{0}{6}
    
	\end{axis}
\end{tikzpicture}
    \subcaption{Per participant}
    \label{fig:boxplot-informal-model-kind-per-participant}
    \end{subfigure}
    \caption{\sosym{Number of different model kinds informally depicted by participants (a), where the * indicates answers from the free-form field, and a box plot highlighting the distribution of model kinds used per participant (b).}}
    \label{fig:hist-informal-model-kinds}
\end{figure}

\subsubsection{RQ3.1: For which purpose are informal models used?}
To investigate how informal models are employed in practice,
our first question asked participants to rate how frequent they emply informal models for each purpose.
\autoref{fig:rq3_a_boxplot} plots their answers indicating that informal models are often used for communication and documentation.
Participants who chose ''rarely'' or ''not at all'' are considered outliers.
Besides that, informal models were often used for the design and requirements analysis of \dts, although 25 \% of the participants never or rarely use informal models for these purposes.
In addition, three participants entered other purposes stating that
informal models were used for \textit{designing methods for continuous improvements or for updates}, for \textit{modeling development process models}, for \textit{evaluating use cases} or for \textit{developing software architecture}. Although we would consider the latter two to correspond to the \textit{design of \dts}.

\subsubsection{RQ3.2: For which phases of automotive development are informal models employed?}
To survey in which phase of development informal models are typically utilized, participants rated the frequency they use informal models in each phase.
\autoref{fig:rq3_b_boxplot} shows that informal models are often used in the \textit{conception} and \textit{development} phase.
In contrast, the median frequency during the \textit{production}, \textit{usage \& support} or \textit{retirement} phase is \textit{rarely used}, \revone{while the frequency in conception or development phase is \textit{very often} respectively \textit{rather often}}.

\begin{figure}
    \centering
    \begin{tikzpicture}[scale=0.8]
	\begin{axis}[width=1.2\linewidth, height=4cm, xmin=-0, xmax=6,
		   ymin=1, ymax=4,
		   ybar, ymajorgrids, enlargelimits=0.1,
		   legend style={anchor=north east,legend columns=6},
		   scaled ticks=false, tick label style={/pgf/number format/fixed},
		   xtick={0,1,...,6}, 
          xticklabels={\parbox{10em}{(1) Not reused, as not saved},\parbox{10em}{(2) Not reused, although saved},\parbox{10em}{(3) Photographed and reused},\parbox{10em}{(4) Photographed, but not reused},\parbox{10em}{(5) Manually transcribing is time-consuming}, \parbox{10em}{(6) Manually transcribing was useful}, \parbox{10em}{(7) Finding retained models is hard}},
          x tick label style={rotate=90,anchor=east},
          ytick={1,2,3,4},
		   yticklabels={-~-,-,+,++},
		  ]

    \boxplot[red]{0}{2}{2}{3}{1}{4}
    \boxplot[red]{1}{2}{2}{3}{1}{4}
    \boxplot[red]{2}{3}{2}{3}{1}{4}
    \boxplot[red]{3}{2}{2}{3}{1}{4}
    \boxplot[red]{4}{3.5}{3}{4}{2}{4}
    \boxplot[red]{5}{3}{3}{3}{3}{3}
    \addplot[only marks,mark=o,fill=none,red,forget plot] coordinates {(5,1) (5,2) (5,4)};
    \boxplot[red]{6}{2}{2}{3}{1}{4}
        
	\end{axis}
\end{tikzpicture}
    \caption{\sosym{Boxplot of the agreement with statements on the use and reuse of informal models. The agreemant ranges from \textit{disagree at all} (-{}-), via \textit{rather disagree} (-) and \textit{rather agree} (+) to \textit{agree at all}~(+{}+).}}
    \label{fig:boxplot-informal-affirmation-reuse}
\end{figure}
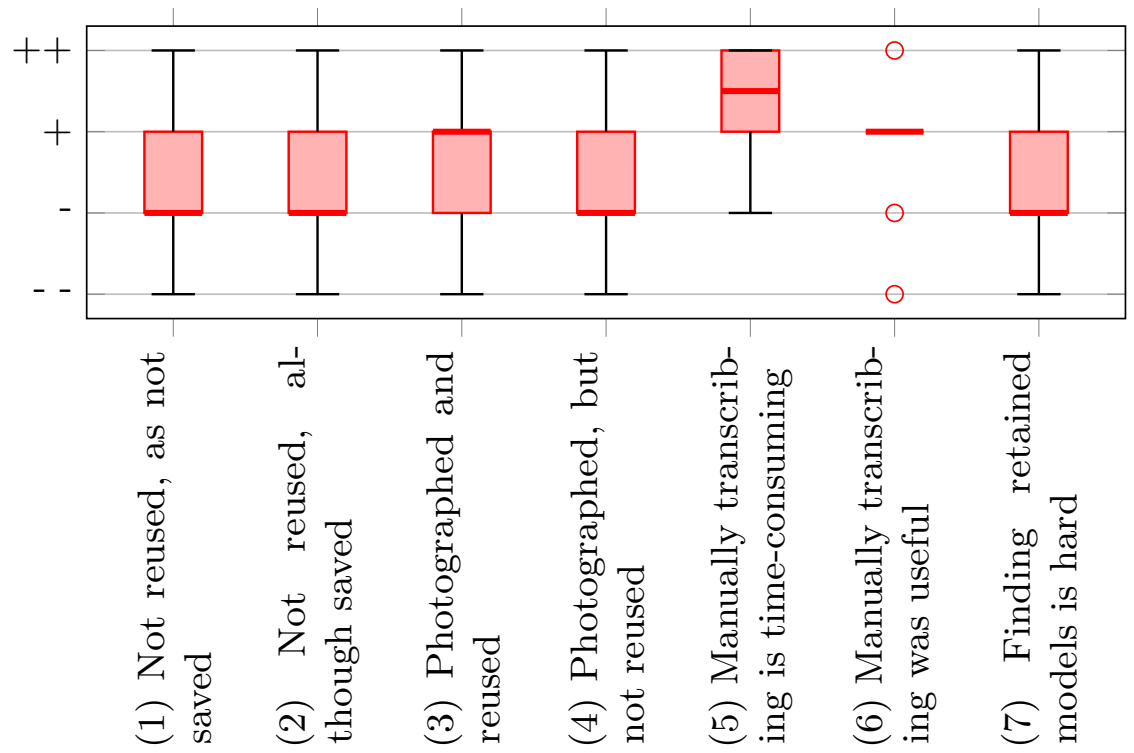

\subsubsection{RQ3.3: How are informal models persisted?}
To gain insights into how informal models are retained, participants rated how often an informal model was created digitally, transcirbed to a formal notation, stored as photograph and retained on the white board.
The boxplots in \autoref{fig:rq3_c_boxplot} show that 75\% of the participants create informal models digitally, whereas only outliers rarely or never create informal models digitally.
Moreover,  50\% of the participants rather often stored photographs of informal models. Still 25\% rather rarely or never photograph informal models.
In regard to transcribing informal models and retaining them on whiteboards, the answers of participants ranged from very often to never, whereas the median answer was rather rarely.


\subsubsection{RQ3.4: Which kinds of models are informally designed?}
To collect the kinds of informal models participants use, we asked participants to select any number of model kinds from a subset of seven model kinds or name an additional model kind.
\autoref{fig:hist-informal-model-kind} summarizes the answers, whereas \textit{activity diagrams} and \textit{state diagrams} were selected most often, closely followed by \textit{business process models} and \textit{development process models}.
These three model kinds describe a system's or process' behavior.
All other model kinds were selected less often.
Additionally, three participants named \textit{software architecture models} (or the synonym component models) and two \textit{deployment models} as additional models.
Besides that, \autoref{fig:boxplot-informal-model-kind-per-participant} indicates that the number of different model kinds selected per participant is three, ranging from none to six. Only 6 participants did not select or named any model kind.

\begin{figure}[tb]
    \centering
    \includegraphics[width=\linewidth]{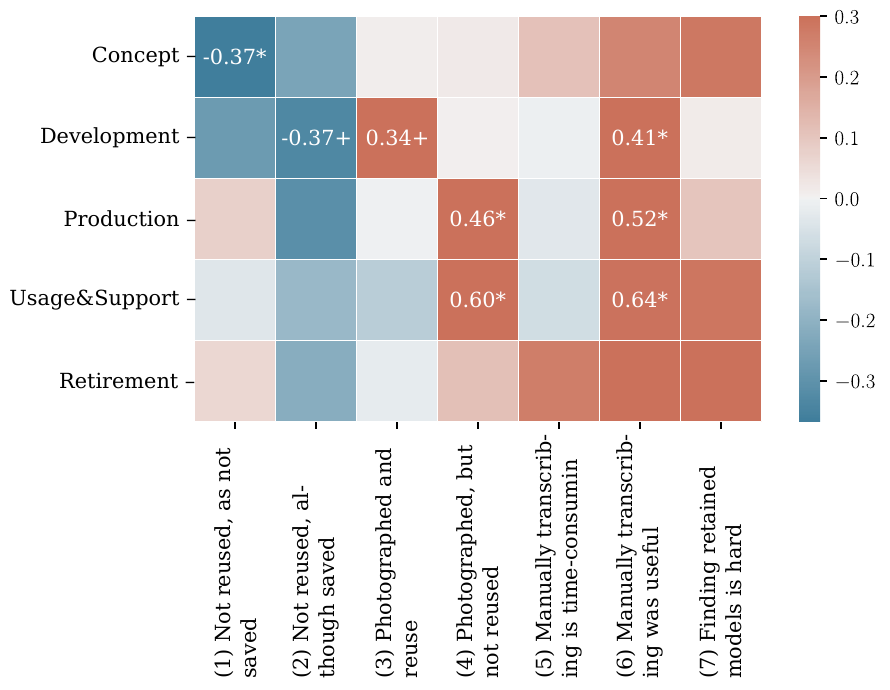}
    \caption{\revone{Heatmap of the correlations between how frequently informal models are used in each phase of development and how informal models are used and reused. Hue indicates a positive or negative correlation. Strong significance ($p<0.05$) is indicated with a \texttt{*} and weak significance ($p<0.10$) with~a~\texttt{+}.}}
    \label{fig:heatmap-informal-phase-persisting}
\end{figure}

\begin{figure}[tb]
    \centering
    \includegraphics[width=\linewidth]{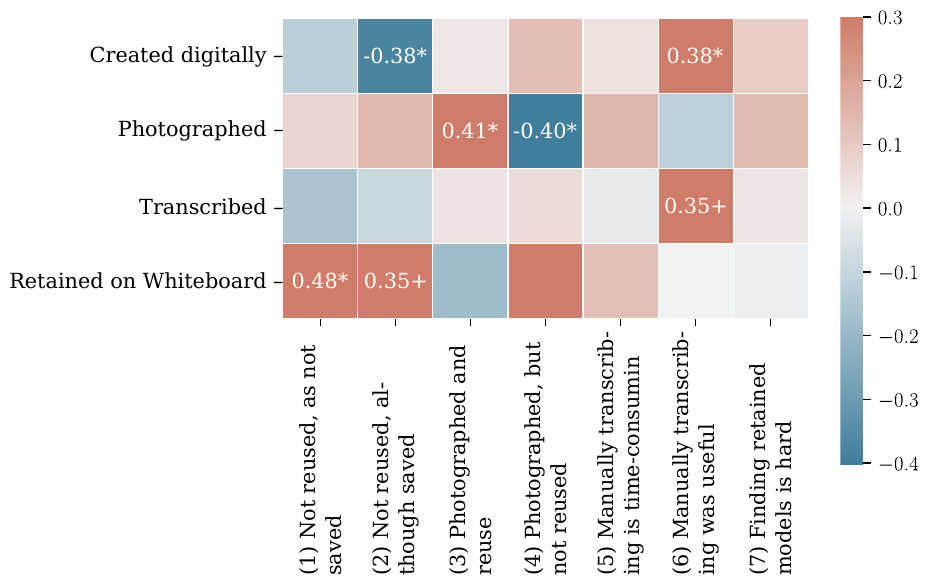}
    \caption{\revone{Heatmap of the correlations between the way informal models are persisted and how they are used and reused. Hue indicates a positive or negative correlation. Strong significance ($p<0.05$) is indicated with a \texttt{*} and weak significance ($p<0.10$) with a \texttt{+}.}}
    \label{fig:heatmap-informal-phase-persisting-purpose}
\end{figure}

\subsubsection{RQ3.5: How are informal models re-/used?}\label{sec:results-rq3.5}
Finally, to investigate how participants use and reuse informal models in practice, we asked the participants to rate the agreement with seven statements on how informal models are stored and reused (cf.~\autoref{sec:methodology-informal}).
In \autoref{fig:boxplot-informal-affirmation-reuse}, we collected the results.
The results indicate that the median \textit{rather disagrees} with statements (1), (2), (4), and (7). Thus, disagreeing with statements including not reusing stored informal models.
In contrast, the median of participants \textit{rather agreed} with Statement~(3) indicating that informal models are photographed and reused.
Moreover, the median of the agreement with Statement (5) regarding the time consumption of manually transcribing informal models is between rather agree and agree at all, whereas no participants completely disagreed.
Furthermore, all participants (except outliers) rather agree with Statement~(6) regarding the usefulness of manually transcribing informal models.


Besides agreement with the seven statements, we also asked participants to describe obstacles hindering the reuse of informal models.
Four participants provided such descriptions. We paraphrase their descriptions as:
\begin{itemize}
    \item \textit{Versioning is another problem},
    \item \textit{Lack of analytical ability and potential misunderstandings},
    \item \textit{(Partly) incorrect assumptions made for informal models}, and
    \item \textit{Learning of rarely used modeling tool to transcribe informal model}.
\end{itemize}

\revtwo{In addition to the presented results, we conducted a correlation analysis to identify relations between how frequently different kinds of informal models are employed in each phase of development as well as the way informal models are persisted and (re)used.}
In \autoref{fig:heatmap-informal-phase-persisting}, we depict the correlations between how often informal models are used in a particular phase of development and the agreement with the statements of reuse. 
\revtwo{
Here, we observe a statistically significant correlation between the agreement with Statement~(6), i.e., that manual transcription of informal models is useful, and the frequency of their use in \textit{development}~\revone{($p=.027$)}, \textit{production}~\revone{($p=.006$)} or \textit{usage \& support}~\revone{($p=.0005$)}.
Moreover, we see a statistically significant correlation between Statement~(4) regarding not reusing photographed informal models and the frequency of their use in \textit{production}~\revone{($p=.026$)} and \textit{usage \& support}~\revone{($p=.0023$)}.
Finally, we found a statistically significant negative correlation between how often informal models are used in the conception phase and agreement with Statement (1)~\revone{($p=.046$)}, i.e., that informal models are not reused as they are not persisted.
Besides that, we found correlations between the frequency of using informal models in the development phase and Statement~(2)~\revone{($p=.049$)} and Statement~(3)~\revone{($p=.097$)}. These, however, have a weak statistical significance.}

\revtwo{
The second correlation analysis, shown in \autoref{fig:heatmap-informal-phase-persisting-purpose}, illustrates correlations between the frequency of persisting informal models in a particular way and the agreement to statements on re-/use.
We identified a statistically significant negative correlation between the frequency of digitally created informal models and agreement with Statement~(2)~\revone{($p=.032$)}, i.e., that informal models are not reused, although they were saved, as well as a positive correlation with Statement~(6)~\revone{($p=.037$)}, i.e., that manually transcribing an informal model was useful.
Similarly, we found a statistically significant positive correlation between the frequency of informal models retained as a photograph and agreement with Statement~(3)~\revone{($p=.028$)}, i.e., that informal models are photographed and reused, as well as a negative correlation with Statement~(4)~\revone{($p=.034$)}, i.e., not reusing photographs of informal models.
Moreover, we identified a statistically significant correlation between how often informal models are retained on a whiteboard and agreement with Statement~(1)~\revone{($p=.008$)}, i.e., not reusing the informal model as it was not saved.
In addition, we found two correlations of weak significance:
The first, between the frequency of informal models being retained on white boards and agreement with Statement (2) not reusing informal models~\revone{($p=.056$)}.
The second, between how often are transcribed using an appropriate tool and Statement~(6) stating the usefulness of manual transcription of informal models~\revone{($p=.0.06$)}.
}
}
\section{Discussion}
\label{sec:discussion}
In this section, we discuss our findings and answer our research questions.
Moreover, we discuss how we avoided threats to validity.
\revtwo{Furthermore, we reassess the presented correlations to address potential multiple correlation biases.}

\subsection{RQ1: How is the DT understood in the automotive industry?}

\subsubsection{RQ1.1: For which phases of automotive development are \dts important?}
Concerning the relevance of \dts in different phases of automotive development (cf.~\autoref{fig:rq1_a_boxplot}) our participants from the automotive domain agreed that the \dt is relevant in all phases of development.
Most interestingly, the \dt is considered \textit{very important} in the \textit{usage \& support} phase.
This indicates that the \dt is considered to be most useful when the \cps is already in use.
This might be related to the \dt's ability to bridge the gap between the physical and the digital world and to provide insights into the physical system's behavior and performance.
The second most important phase is the \textit{development} phase, where the \dt could prevent early errors and provide insights on the vehicle that would otherwise be only available after production. For instance, if the \dt of the vehicle comprises a virtual representation, this could be used to simulate certain vehicle functionalities where otherwise, a physical prototype would be necessary in this phase of development.

\subsubsection{RQ1.2: What are desired properties of \dts?}
Concerning the properties of \dts (cf.~\autoref{fig:rq1_b_boxplot}) our participants from the automotive domain agreed on all provided properties.
Nevertheless, it is important to see the differences regarding their importance.
In this regard, the data shows that the \textit{as-operated} property is considered very important.
In contrast to that, the properties \textit{as-manufactured} and \textit{as-designed} are considered important.
This shows that \dts are especially interesting for the operational state of \cpss.
The fact that the property \textit{virtual representation} is considered very important, too, aligns with the answers for RQ1.1, where the \textit{development} phase is considered one of the most important ones. Here, simulating the vehicle in development can be very helpful.
The other very important properties \textit{collects data} and \textit{bi-directional} align with the automotive development phases \textit{usage \& support} and \textit{development}.
These provide valuable insights into the actual \cps.

\subsubsection{RQ1.3: What are desired purposes of using \dts?}
Concerning the purposes of \dts (cf.~\autoref{fig:rq1_c_boxplot}) our, participants from the automotive domain agreed that \textit{prediction}, \textit{monitoring}, and \textit{validation} are the most important purposes. This indicates that the purpose of using the \dt is focused on collecting data to observe and predict the behavior of a \cps, but not actively controlling or interacting with the \cps.
The purposes of \textit{homologation} and \textit{coordination} of multiple twinned systems had the full range of answers, from not important to very important.
Because some experts of the consortium actively work on improving homologation and fleet management, this might introduce some bias toward the importance of the topic, while other participants did not consider these purposes important.

\subsubsection{RQ1.4: How do these purposes change concerning different phases of automotive development?}
For the correlation between the answers for relevant purposes and development phases of \dts (cf.~\autoref{fig:rq1_d_heatmap}), we found that the \dt is considered useful in all automotive development phases. 
There is a strong correlation between the purpose \textit{monitoring} and the post-production phases\textit{usage \& support} and \textit{retirement}. This indicates that the \dt should monitor the running system for instance to recognize erroneous behavior early and to gain knowledge from the \cps when it is retired, for instance, to analyze historical data for a new product generation.
Since \textit{validation} is required and important especially when \textit{conceptualizing} a product, its \textit{development} and its \textit{production}, there is a strong correlation between this purpose and these three development phases.
The purpose of \textit{homologation} strongly correlates with the \textit{concept} phase. This indicates that automotive experts want to know as early as in the concept phase whether the product is going to be accepted or what might prevent it from becoming accepted by the authorities. 
Our findings indicate that the purpose \textit{coordination} is relevant for the \textit{concept} phase too. This enables the design concept to already interact with other systems. For \textit{retirement}, it seems important to inform or to know the other systems the \cps is coordinated with.
The purpose \textit{prediction} only correlates with the \textit{usage \& support} of the \cps.
This indicates that the prediction is mostly useful during the usage of the \cps, \eg to predict future behavior or maintenance cycles.
Concerning the purpose of \textit{process optimization}, the answers mostly correlated with the phase \textit{production}.
This indicates that the participants mostly thought of the production process when aiming for process optimization.
In fact, all phases include processes to be optimized.
\textit{Risk assessment} strongly correlates with the phases of \textit{production}, \textit{usage \& support} and \textit{retirement}. As such, this indicates that the automotive industry experts consider these phases as the most risk full, \eg production errors or outages, malfunctions during the usage of the \cps or expensive retirement costs.

\vspace{5pt}
\begin{mdframed}[style=rqanswer]
{\bf Answer to RQ1:}
In the automotive industry, the \dt is considered most useful in the \textit{usage \& support} phase of automotive development. This is supported by the fact that most participants considered that the \dt should describe the \cps \textit{as-operated}. To this end, the most important purposes of the twins are the monitoring of data, performing validations and predicting future behavior.
However, the purposes of the \dt depend on the phase of automotive development.
\end{mdframed}

\subsection{RQ2: Which modeling languages and modeling tools are currently employed in the automotive industry?}


\subsubsection{RQ2.1: Which kinds of models are important in automotive development?}
Regarding the kinds of models relevant to the automotive domain, we found that besides \textit{simulation models}, \textit{source code} and \textit{business process models} were the kinds used by most participants.
Notably, though, we found that \textit{deployment models}, previously not considered by us, were employed by some participants.
While this might be in part explained by the distribution of professional backgrounds, we insist that this still highlights the variety of models employed in the automotive domain regardless of the total number of mentions. More importantly, our findings indicate that several model kinds are relevant for each working area and most participants work with two to four different kinds of models.
Thus, for a \dt used during automotive development, it must be able to include and relate multiple different kinds of models ranging from structural, \eg \textit{CAD models}, via behavioral, \eg \textit{business process models}, to quality models, \eg \textit{variability models}.

\subsubsection{RQ2.2: How important is each model kind for the phases of automotive development?}
As the \dt is supposed to support all phases of development, we especially asked for the importance of each model kind for each phase of automotive development.
Although we initially assumed that many model kinds would be considered important in one or two phases, we found that most model kinds are considered important for at least three phases, except \textit{CAD models}.
Moreover, we were surprised to find that \textit{data schemes}, \textit{business process models}, \textit{development processes} and \textit{variability models} are considered important throughout all phases of automotive development.
This indicates that \dts in the automotive domain should consider the management of \textit{data schemas}, \textit{business process models}, as well as \textit{variability models}.
In particular, \dts must handle different versions and variants of automotive.
From the participant's answers, we conclude that all model kinds are relevant for all phases of automotive development.
While some are more important in a phase than others, we argue that all model kinds should be supported by a \dt, regardless of whether it is used in one phase or multiple phases.

\subsubsection{RQ2.3: Which tools are used to create and maintain these models?}
Finally, we were interested in the kinds of tools and development environments employed in the automotive domain and whether they already support multiple different model kinds.
Surprisingly, we found that only a minority of tools integrate two (or more) model types, such that both models can be created, and provisioned, and their consistency be ensured.
\textit{PREEvision} is a model-driven development environment\cite{schauffele2016} supporting multiple layers of abstraction from requirements down to the car's electrical wiring, the tool landscape is characterized by isolated tools tailored to one particular model kind.
Consequently, to introduce a \dt into the automotive domain would require connecting to a large variety of tools, as participants would want to continue using the existing tools.
Although there are established standards in the automotive domain, e.g., AUTOSAR~\cite{fuerst2016}, used for the exchange of structural and to some extent behavioral information about a specific automotive, for many model kinds there is simply no standard.
We conclude that, for a \dt to be successful in the automotive domain, it must be able to connect to and/or interact with many different established, commercial tools and development environments.
Granted, this will be a major challenge for the modeling community.

\vspace{5pt}
\begin{mdframed}[style=rqanswer]
{\bf Answer to RQ2:}
\textit{Simulation models}, \textit{source code}, and \textit{business process models} are the most important models used. However, in all phases of automotive development, models are relevant and used in practice. Concerning tools, they are mostly limited to one modeling language. Since all models are relevant in the different phases of automotive development, the tools that allow their creation are too. Therefore, we conclude that establishing a \dt in the automotive domain requires it to integrate and interact with many different modeling languages and tools.
\end{mdframed}

\sosym{
\subsection{RQ3: How relevant are informal models in the automotive industry?}

\subsubsection{RQ3.1: For which purpose are informal models used?}
To understand why and how informal models are used, stored and reused, we first investigated which purpose informal models fulfill.
We found that they often used for communication, documentation, the design of \dts as well as requirements analysis.
However, it is evident that most participants use informal models for communication and documentation.
This is in line with findings about software development processes \cite{cherubini_lets_2007}.
Moreover, as participants named additional purposes, e.g., \textit{modeling of development processes}, we conclude that these purposes are not exhaustive.

\subsubsection{RQ3.2: For which phases of automotive development are informal models employed?}
Regarding the phases of automotive development, we found that informal models are most often utilized during the conception and development phase. 
While we observe that informal models were also regularly employed in other phases of automative development, most particpants rather rarely employ them during the production, usage \& support, and retirment phase.
We argue that this is due to the way informal model can help participants to cost efficiently discuss and capture early design decisions.
Thus, we see a chance in integrating informal models into digital twins to support the early development phases and to make documented design decisions available the whole life cycle of the \dt.

\subsubsection{RQ3.3: How are informal models persisted?}
As informal models were mostly employed for communication and documentation, we surveyed how participants recorded them for later examination or revision.
Surprisingly, we found that informal models were most frequently created using digital tools, e.g., PowerPoint\texttrademark, and thus could be stored directly.
Moreover, photographs of informal models sketched on flipcharts or whiteboards were still the second frequent option for persisting informal models.
Although this observation is in contrast to past research on the use of informal models in software development~\cite{baltes_sketches_2014} showing that most informal models are created using pen and paper or whiteboards, we argue this is due to the increased frequency of online meetings of engineers working in different locations. Granted, this could be a temporary shift following the COVID-19 pandemic.
Notably though, some participants still manually transcribe informal models or simply keep them on the flipchart or whiteboard, although most participants \textit{rather rarely} do either.
In conclusion, \dts should be able to support the integration of digitally created informal models as well as digital images (e.g., a photograph or trace of a digital pen) of informal models.
While the former could be easy, the latter would require image recognition pipelines, like \cite{KoziolekKoziolek2024_1000175780}, tailored to the different model kinds typically sketched.

\subsubsection{RQ3.4: Which kinds of models are informally designed?}
To investigate which model kinds are actually sketched by participants, we tasked them to select or name those model kinds.
We expected that participants would name equally many structural model kinds, such as \textit{E/E architecture models} or \textit{data schemas}, as behavioral model kinds, such as \textit{activity diagrams} or \textit{business process models}.
Yet, we found that behavioral model kinds were named significantly more often, i.e., \textit{activity diagrams} and \textit{state diagrams} followed by \textit{business process models}. 
We conclude that this might be due to the emphasis on an automotive's behavior or business processes around the automotive during the \textit{conception} phase.
Notably, the number of different model kinds sketched by participants (\autoref{fig:boxplot-informal-model-kind-per-participant}) is comparable to the number of different model kinds designed (\autoref{fig:boxplot-model-kind-per-participant}), although some participants stated to never draw informal models.
In sum, a wide variety of model kinds are sketched, even though behavioral model kinds are more common.
In turn, developers of \dts should start by supporting the integration of the most common behavioral model kinds sketched by practitioners in the automotive industry, i.e., \textit{activity diagrams}, \textit{state diagrams}, and \textit{business process models}.

\subsubsection{RQ3.5: How are informal models re-/used?}
In our final question, we investigated how informal models are reused (after they were drawn and persisted) by means of seven statements.
While most participants rather disagree with statements regarding not reusing informal models, most agree with the statement that the \textit{manual transcription of an informal model is time-consuming yet is considered useful}. Additionally, a participant described the manual transcription as a \textit{hurdle}, if the modeling tool is rarely used.
In sum, we see a need for the automatic transcription of informal models, e.g., by means of image recognition.
Besides that, from the agreement with Statement~(3) and disagreement with Statement~(4), we conclude that \textit{if an informal model is photographed, it is more likely to be reused}.

From the correlations between the phase of development and the presented statements on reuse (cf.~\autoref{fig:heatmap-informal-phase-persisting}), we conclude that participants who frequently employ informal models in the conception phase usually reuse and save their models.
In addition, we conclude that participants employing informal models in later development phases (except retirement) would not reuse photographed models but rather a transcribed model.
Likewise, from the correlations between the way informal models are stored and the statements on reuse (cf.~\autoref{fig:heatmap-informal-phase-persisting-purpose}), we entail that informal models only retained on a whiteboard or flipchart are rarely reused.
Surprisingly, whenever participants have created informal models digitally, they are more likely to reuse and transcribe them.
Moreover, we imply that participants who often transcribed informal models also used a photograph of an informal model as reference, yet some do not consider this task to be useful.
\revone{An experienced reader might interject that it is a common approach to sketch the most important concepts and relations on a whiteboard or in PowerPoint\texttrademark leaving out unnecessary details. Afterward the sketch will be modeled and thus formalized with an appropriate tool, which complains about missing details or incomplete models.
While we agree that this is a very common practice, we argue that this is a shortcoming of modeling tools not allowing for incomplete models or sketches, to be later completed.
Because a sketch persisted in the modeling tool, albeit incomplete, it can be synchronized and retrieved within the digital twin.}


\vspace{5pt}
\begin{mdframed}[style=rqanswer]
{\bf Answer to RQ3:}
\sosym{The results of the survey indicate that informal models are frequently used in the automotive industry, in particular during the \textit{conception} and \textit{development} phase.
Informal models are mostly used for the \textit{communication} and \textit{documentation} of design decisions. Surprisingly, most participants frequently \textit{create informal models digitally}, e.g., using PowerPoint\texttrademark, although they admit that \textit{manually transcribing models is time-consuming}.
Last but not least, participants sketch more behavioral models, most importantly, \textit{activity diagrams} or \textit{state diagrams}, then structural models.}
\end{mdframed}
}

\begin{figure}
    \centering
    \begin{tikzpicture}[scale=0.8]
	\begin{axis}[width=\linewidth, height=4cm, xmin=-0.5, xmax=0.5,
		   ymin=1, ymax=4,
		   ybar, ymajorgrids, enlargelimits=0.1,
		   legend style={anchor=north east,legend columns=6},
		   scaled ticks=false, tick label style={/pgf/number format/fixed},
		   xtick={0}, 
          xticklabels={Agreement to our definition of \dts},
          ytick={1,2,3,4},
		   yticklabels={-~-,-,+,++},
		  ]

    \boxplot[blue]{0}{4}{3}{4}{2}{4}
    \addplot[only marks,mark=o,fill=none,blue,forget plot] coordinates {(0,1)};
        
	\end{axis}
\end{tikzpicture}
    \caption{\sosym{Boxplot agreemant with the definition: 
    \emph{<<A digital twin is a software comprising models and services to purposefully represent and manipulate its original system across its complete lifecycle.>>}. The agreemant ranges from \textit{disagree at all} (-{}-), via \textit{rather disagree} (-) and \textit{rather agree} (+) to \textit{agree at all}~(+{}+).}}
    \label{fig:boxplot-dt-definition}
\end{figure}
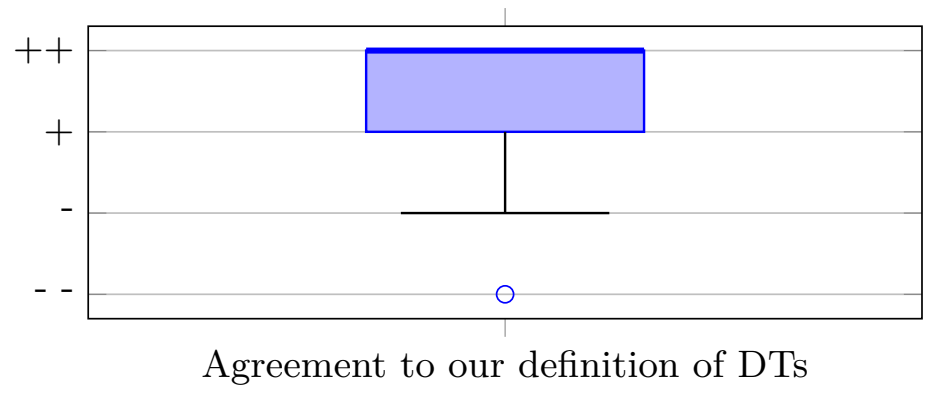

\sosym{
\subsection{Agreement to Our Definition of Digital Twins}
As the last question, we asked participants to rate their agreement with our definition of \dts (cf.~\autoref{sec:background_dt}).
\autoref{fig:boxplot-dt-definition} shows that $75\%$ of participants agree with the definition.
Notably though, only one participant disagreed completely.
We argue that this overall agreement stems from the fact that the definition was derived from the literature and has been discussed with experts in the field.
Although we concede that this might be caused by the questions and aspects of our questionaire.
However, the disagreement of one participant might indicate that our definition is not yet perfect and needs further refinement or that there are different opinions on the definition of digital twins in the automotive domain.
Ultimately, we consider our definition to be a good starting point for further research on a common definition of \dts in the automotive domain. 
}

\revtwo{
\subsection{Bonferroni Correction of the Presented Correlations}\label{sec:bonferroni}

Although the indicated correlations were meant to present an initial analysis, we concede that the significances of the presented correlations for RQ1.4 in~\autoref{fig:rq1_d_heatmap} as well as for RQ3.5 in~\autoref{fig:heatmap-informal-phase-persisting} and \autoref{fig:heatmap-informal-phase-persisting-purpose} are biased due to multiple correlations being considered.
We applied the \emph{Bonferroni correction}~\cite{dunn1961multiple} to further lower the significance levels wrt.~the number of correlations tested.
In regard to RQ1.4 (cf.~\autoref{fig:rq1_d_heatmap}), we tested $m=35$ correlations, lowering the significance level for a strong and weak significant correlation to $p<.05/35\approx .00143$ and $p<.1/35\approx .00286$, respectively. None of the reported $p$-values in \autoref{sec:results-rq1.4} crossed these levels.
For RQ3.5 (cf.~\autoref{sec:results-rq3.5}), the correction of the correlation analysis presented in~\autoref{fig:heatmap-informal-phase-persisting-purpose} with $m=28$ requires a significance level of $p<.00179$ for a strong correlation and $p<.00357$ for a weak correlation. Again, the reported $p$-values did not cross these thresholds.
In contrast, the correlation analysis presented in~\autoref{fig:heatmap-informal-phase-persisting} (with $m=35$) still indicates a strong correlation ($p<.00143$) between Statement~(6) that manual transcription of informal models is useful when frequently employed for \textit{usage \& support}~\revone{($p=.0005$)}. Likewise, a weak correlation $(p<.0028)$ is maintained between Statement~(4) that photographs are not reused when they are taken during \textit{usage \& support}~\revone{($p=.0023$)}.
Consequently, our survey only provided initial evidence for RQ1.4 and RQ3.5, yet most results are not significant under the Bonferroni correction. Please note that this does not impact the other 10 research questions (RQ1.1-1.3, RQ2.1-2.3, \& RQ3.1-3.4).
}

\subsection{Threats to Validity}
Our survey is subject to the four basic types of validity threats according to~\cite{wohlin2012experimentation} are construct validity (research design), internal validity (data extraction), external validity (generalizability), and conclusion validity (reliability).

\textit{Construct validity} refers to how well the study represents what the researchers intended it to represent in the study construction.
In this study, we relied solely on a questionnaire to measure the definition of \dts and the models and tools employed by the German automotive industry.
To mitigate this issue, we did two pre-studies and discussed the questionnaire content with experts in the field of modeling and \dts.
Another threat to construct validity could arise from discretizing the importance of the different properties of \dts, modeling languages, and used tools using balanced four-point Likert scales.
This, however, has been investigated and identified as not compromising construct validity~\cite{ray1982construct}.
Another threat to construct validity could originate from too many questions overwhelming the participants and hindering them from completing the questionnaire.
To mitigate this threat, we showed questions that depended on the agreement with another question only on the agreement, thereby presenting every participant with the most minimal version of the questionnaire.

A threat to \textit{internal validity} could arise from the different understandings of what a \dt is and the fact that the questionnaire was filled out online without supervision. 
To mitigate this effect, we performed multiple pilot studies before sending out the survey to the participants.
Since our survey targeted German automotive experts within a German government-funded project, we designed the survey in German language.
To mitigate this issue, we provide the original questionnaire in German online, together with the original response data and a translated version of the data in English.
Although the number of participants was small and their answers might not exhibit a normal distribution, according to Geoff~\cite{norman2010likert}, the Pearson correlation coefficient and Student's T-test are still applicable in our setting.
\revtwo{As we were interested in initial correlations, we did not apply means to correct for false-positives of the found significances. To remedy this, in~\autoref{sec:bonferroni}, we reassessed the three correlation analyses using the suggested Bonferoni correction. As a result, the presented correlations should be considered as initial results requiring further surveys and questionnaires to provide conclusive evidence.}

Threats to \textit{external validity} indicate to which extent the results of our study can be generalized.
One threat to external validity is that most of our 55 of 115 participants are computer scientists and thereby the answers to our survey could be biased towards their view on the definition of \dts and the importance of certain models and modeling tools.
\add{Another threat to external validity is that we only surveyed members of the SofDCar research project that are representatives of the German automotive industry and, thus, could prevent our results from being generalizable to the automotive industry of other countries. }
\revone{Furthermore, we did not collect information on whether respondents were from academia or industry. 
As a result, we cannot rule out potential bias introduced by a higher proportion of academic participants.}

To mitigate both threats, we call for replication to extend and generalize the results of the study to other disciplines that leverage \dts and models in practice.

Threats to \textit{conclusion validity} entail avoiding wrong conclusions and permitting replicability.
Regarding the former,
because the answers to the questions were selected by the authors of this paper and could introduce some bias towards desired properties of \dts or the importance of modeling language or tools given by us\revone{, \eg by not including deployment models.}

To mitigate this issue, we always included an open-answer field for input that was not initially part of the survey.
For the study’s replicability, we detailed the complete research method in \autoref{sec:methodology}, which enables replicating the methodology of this questionnaire survey. In addition, we provide the data extraction results and the original survey online.\footref{reppack}

\section{Related Studies}
\label{sec:relatedwork}

This section presents related studies that survey the properties, capabilities, and goals of \dts, modeling languages and tools that are employed in automotive development, and studies that investigate informal modeling practices.
\revtwo{For each group of related studies, we provide a structured comparison table that classifies how each related finding relates to our survey results: \textit{Confirmed} findings are corroborated by our data, \textit{Extended} findings are covered by our survey with additional depth or scope, and \textit{Contrast} marks findings that differ from, or are out of scope of, our study.}

\subsection{Surveys on Digital Twins}
\dts are relevant in many domains and, as such, objective to current research activities.
A systematic mapping study on software engineering for \dts~\cite{DJR+22} investigated the purposes and life cycle phases considered relevant for \dts in the current literature.
They concluded that the main purpose of \dts is behavior optimization, monitoring, and prediction, especially in the runtime of the \cps.
These observations align with the responses from our survey where monitoring, prediction, and optimization together with validation are the main purposes of employing \dts.
Besides, a systematic mapping study investigated the state of the art in employing models in the development of digital twins in research~\cite{DBLP:journals/sosym/LehnerZPSSWW25}.
In comparison to this work, we focused on the automotive domain and surveyed experts from industry.
Another literature survey focused on the technical aspects of realizing \dts~\cite{DBLP:journals/jim/LimZC20} and identified four essential categories of technologies that are required to work together that are communication, representation, computation, and microservices.
Since the aim of the SofDCar research project is building a \dt from scratch the technologies to build a \dt were not the scope of our questionnaire.
A survey among 22 participants from the production planning department of a major automobile manufacturer~\cite{biesinger2019survey} showed that \dts increase the quality and lower the complexity in automotive production planning.
The added value of \dts in production is also reflected in our study where our participants identified production as the second most important phase of development (\cf \autoref{fig:rq1_a_boxplot}).
Another systematic literature review on \dts in automotive application~\cite{deng2023systematic} differentiates between the purposes of a \dt for manufacturing, which can monitor and optimize the production of an automotive, and a \dt for the automotive in operation, where the \dt can predict, monitor and optimize the energy consumption of an automotive.
Our survey also investigated the question of which purposes of a \dt is relevant in which phase of automotive development (\cf \autoref{fig:rq1_d_heatmap}).
We identified validation, process optimization and risk assessment to be desired purposes in the production phase and monitoring, prediction, and risk assessment in the usage \& support phase.
\revone{
Another investigation presents challenges for the development of digital twins~\cite{DBLP:journals/software/AbrahaoSMCC25}.
It considers only papers from the 27th International Conference on Model Driven Engineering Languages and Systems and, consequently, limited in scope.
However, it identifies the derivation of UML diagrams from informal models as one challenges, which aligns with our results that providing digital twins with informal models is beneficial.
}
\revtwo{
\autoref{tab:related-dt-surveys} systematically classifies which of these findings are confirmed, extended, or contrasted by our results.
}
\begin{table*}[tp]
\revtwo{

\caption{Comparison of related surveys on \dts with the findings of our survey.}
\label{tab:related-dt-surveys}
\begin{tabularx}{\linewidth}{lXX}
\toprule
\textbf{Study} & \textbf{Key Finding} & \textbf{Alignment with Our Survey} \\
\midrule
Dalibor et al.~\cite{DJR+22}
  & Systematic mapping of SE for \dts: main purposes are behavior optimization, monitoring, and prediction, especially at CPS runtime.
  & \textit{Confirmed}: monitoring, prediction, and validation reach a median of \textit{very important}; process optimization is also rated as an important purpose, aligning with the behavior optimization finding of Dalibor et al. (\cf \autoref{fig:rq1_c_boxplot}). \\
\addlinespace
Lehner et al.~\cite{DBLP:journals/sosym/LehnerZPSSWW25}
  & Surveys model-based \dt development approaches in research literature.
  & \textit{Extended}: we focus on the automotive domain and on practitioners rather than research literature. \\
\addlinespace
Lim et al.~\cite{DBLP:journals/jim/LimZC20}
  & Four essential technology categories for \dts: communication, representation, computation, and microservices.
  & \textit{Out of scope}: technology realization was not the focus of our questionnaire, which targeted purposes and model kinds used in practice. \\
\addlinespace
Biesinger et al.~\cite{biesinger2019survey}
  & \dts increase quality and reduce complexity in automotive production planning (n\,=\,22).
  & \textit{Confirmed}: production is the second most important development phase (\cf \autoref{fig:rq1_a_boxplot}). \\
\addlinespace
Deng et al.~\cite{deng2023systematic}
  & Distinguishes \dt purposes for manufacturing (monitor, optimize production) from \dt-for-vehicle purposes (predict, monitor, optimize energy).
  & \textit{Extended}: our study correlates specific purposes with development phases, \eg process optimization with production and prediction with usage \& support (\cf \autoref{fig:rq1_d_heatmap}). \\
\addlinespace
\revone{Abrah\~{a}o et al.~\cite{DBLP:journals/software/AbrahaoSMCC25}}
  & \revone{Identifies challenges in \dt development; highlights derivation of UML diagrams from informal models.}
  & \revone{\textit{Confirmed}: providing \dts with informal models is beneficial, as also reflected in our RQ3 results.} \\
\bottomrule
\end{tabularx}
}
\end{table*}

\subsection{Surveys on Modeling Languages and Tools in Automotive Development}
A systematic mapping study on modeling languages for Industry~4.0~\cite{WBCW20} identified UML variants and domain-specific languages as the most often applied languages followed by knowledge representations, \eg via data models.
Furthermore, they state that despite validation being an important concern in Industry 4.0, there is little research in this direction.
Compared to our study, the mapping study relies on literature, whereas we questioned experts from the German automotive industry on which modeling languages and tools they use most frequently.
Concerning model-based engineering in the embedded domain, a survey among 113 participants from industry~\cite{DBLP:journals/sosym/LiebelMTLH18} highlighted that models are a central aspect of developing embedded systems. Common languages are MATLAB/Simulink, UML and SysML which are used for simulation and code generation, as well as behavioral and structural consistency checking.
Another survey for modeling and model-driven engineering practices in the embedded industry~\cite{DBLP:journals/jsa/AkdurGD18} showed that informal models are widely used in the embedded industry.

In contrast to these studies, our study focuses on formalized modeling languages, their tools and their relevance for the different phases of development and work areas in the automotive industry.
In particular, we did not investigate the relevance of current research efforts in the automotive domain, such as, automotive reconfiguration~\cite{DBLP:conf/sose/SchindewolfWKGS23}, consistency preservation~\cite{Wittler2023}, handling uncertainty~\cite{Acosta2022}, modelling of variants and versions~\cite{Ananieva2022}, or novel modeling and programming languages~\cite{kuehn2014}. 
Besides surveys covering the state of the art of modeling in general, there exist surveys for model-based security testing approaches~\cite{DBLP:journals/access/SommerKK23}, variability modeling~\cite{DBLP:conf/vamos/BergerRNABCW13}, and requirements modeling~\cite{DBLP:journals/infsof/SepulvedaCC16}.
However, we wanted to provide an overview of common tools and modeling languages in the automotive domain, and surveying the languages and tools in detail was not the scope of our survey and could be potential future work, e.g., employing the assessment grid for intelligent modeling assistance~\cite{mussbacher2020}. 

\revtwo{
\autoref{tab:related-modeling-surveys} classifies which of these findings are confirmed, extended, or contrasted by our results.}
\begin{table*}[tp]
\revtwo{

\caption{Comparison of related surveys on modeling languages and tools with the findings of our survey.}
\label{tab:related-modeling-surveys}
\begin{tabularx}{\linewidth}{lXX}
\toprule
\textbf{Study} & \textbf{Key Finding} & \textbf{Alignment with Our Survey} \\
\midrule
Wortmann et al.~\cite{WBCW20}
  & Systematic mapping for Industry~4.0: UML variants and DSLs most applied; validation underrepresented in research despite being an important concern.
  & \textit{Extended}: rather than mapping research literature, we survey industry practitioners and identify the most prevalent model kinds (source code, simulation models, business process models) and tools (\eg \textsf{MATLAB/Simulink}, \textsf{Enterprise Architect}) in the automotive domain. \\
\addlinespace
Liebel et al.~\cite{DBLP:journals/sosym/LiebelMTLH18}
  & Survey of 113 embedded-systems practitioners: models are central; MATLAB/Simulink, UML, and SysML used for simulation, code generation, and consistency checking.
  & \textit{Confirmed}: MATLAB/Simulink is the most frequently named tool; we extend this to automotive-specific tools and to their relevance per development phase (\cf \autoref{fig:sankey-models}). \\
\addlinespace
Akdur et al.~\cite{DBLP:journals/jsa/AkdurGD18}
  & Survey in the embedded industry: informal models are widely used.
  & \textit{Confirmed}: informal models are frequently used in the automotive industry as well (\cf \autoref{fig:rq3_boxplot}). \\
\bottomrule
\end{tabularx}
}

\end{table*}

\sosym{
\subsection{Studies on Informal Modeling Practices}
Numerous studies have examined the use of informal models, sketches, and diagrams in software development.
Cherubini et al.~\cite{cherubini_lets_2007} conducted semi-structured interviews and a survey to explore how software developers use diagrams.
They found that diagrams are primarily used for communication and are often transient, meaning they are used for a specific task and then discarded.
This finding is consistent with our survey results, where participants discussed how they use informal models.
However, our we also found that informal models are more often created digitally.
Walny et al.~\cite{walny_follow_2011} investigated the lifecycles of diagrams and sketches in software development by conducting multiple interviews with researchers.
They identified various diagram lifecycles, such as brainstorming, drawing a diagram, taking a photo of it, and uploading the photo to a common wiki.
Our survey also addressed how participants persist informal models, reflecting these practices.
Baltes et al.~\cite{baltes_sketches_2014} conducted an exploratory study with three companies and an online survey to understand the role of sketches and diagrams in software engineering.
They found that informal diagrams are valuable resources that document several aspects of the software development process.
This aspect of documentation was also highlighted in our survey, where participants discussed the use of informal models for documentation purposes.
Störrle~\cite{storrle_how_2017} conducted an online survey to determine whether and how conceptual modeling languages are used in software engineering practice.
He observed that conceptual modeling languages like UML are used in practice, but often as informal models (70-79\%).
In contrast to the aforementioned studies, our research focuses on the use of informal models in the context of digital twins and the automotive industry.

\revtwo{
\autoref{tab:related-informal-studies} classifies which of these findings are confirmed, extended, or contrasted by our results.}
\begin{table*}[tp]
\revtwo{
\caption{Comparison of related studies on informal modeling with the findings of our survey.}
\label{tab:related-informal-studies}
\begin{tabularx}{\linewidth}{lXX}
\toprule
\textbf{Study} & \textbf{Key Finding} & \textbf{Alignment with Our Survey} \\
\midrule
Cherubini et al.~\cite{cherubini_lets_2007}
  & Semi-structured interviews and survey with software developers: diagrams are used primarily for communication and are often transient.
  & \textit{Confirmed}: communication and documentation are the top-rated purposes in our survey (\cf \autoref{fig:rq3_a_boxplot}). \textit{Contrast}: automotive practitioners create informal models predominantly digitally, indicating they are less transient in our context. \\
\addlinespace
Walny et al.~\cite{walny_follow_2011}
  & Interviews investigating diagram lifecycles: brainstorming, sketching, photographing, and uploading to a shared wiki.
  & \textit{Confirmed}: photographing is common (50\,\% of participants rather often); however, digital-first creation dominates over whiteboard retention in our survey (\cf \autoref{fig:rq3_c_boxplot}). \\
\addlinespace
Baltes et al.~\cite{baltes_sketches_2014}
  & Exploratory study across three companies and an online survey: informal diagrams are valuable resources for documenting software development.
  & \textit{Confirmed}: documentation is a frequently cited purpose of informal models in our automotive survey (\cf \autoref{fig:rq3_a_boxplot}). \\
\addlinespace
St\"{o}rrle~\cite{storrle_how_2017}
  & Online survey: conceptual modeling languages such as UML are used in practice, mostly as informal models (70--79\,\%).
  & \textit{Contrast}: our study focuses on the automotive domain and the context of \dts; automotive practitioners actively employ dedicated formal modeling tools alongside informal models, which differs from the predominantly informal use of UML-like notations in general software engineering. \\
\bottomrule
\end{tabularx}
}
\end{table*}
}
\section{Conclusion and Future Work}
\label{sec:conclusion}

We have investigated purposes, expected properties, and modeling techniques for \dts in the automotive industry through the lens of German automotive companies and suppliers involved in the SofDCar project. 
Based on a questionnaire survey answered by 96 professional automotive (software) experts, out of which 43 completed all answers to the questionnaire, we found that digital twins are considered most important in the usage and support phase (i.e., once the vehicle exists) and less during other phases. 
Depending on the phase in which the \dt is meant to be used, experts expect it to fulfill different purposes, which might entail that there will be different kinds of digital twins of a single vehicle throughout its life cycle.
The relevance of monitoring during the conception phase of an automotive \dt suggests that they also serve as a foundation for learning and improvement for future vehicles. 
To model digital twins, besides business process models, descriptive mathematical/physical models and predictive simulation models are most important to the participants. 
These are surpassed only by source code, which might hint at a lack of sufficiently expressive modeling languages for some aspects of automotive \dt development. 
Future studies might shed light on the role of source code in the engineering and application of (automotive) \dts.

Overall, we conclude that integrating models into automotive \dts is highly relevant, as expected from the definitions of digital twins. 
Moreover, the modeling languages employed for such \dts differ from the modeling languages employed for \dts in general~\cite{DJR+22}, which, e.g., also includes knowledge representation techniques.
\add{Surprisingly, in contrast to the existing literature, UML and SysML models are not that frequently used.} 
This suggests that automotive \dts indeed are domain-specific, at least, for the employed models. 
Furthermore, as most models are supported by a single tailored modeling tool, to be accepted by practitioners, automotive \dts must be coupled to a variety of existing tools. 
\add{Therefore, future studies could investigate which approaches companies leverage to cope with the integration of these tools and how information across models is persisted, \eg by using knowledge graphs, tool APIs,  exchange formats, or language extension mechanisms~\cite{BHH+17}.}
\revone{We also see the need for follow-up studies. Since respondents rate many aspects, future research should investigate the reasons for these opinions.}
\add{Concerning the adoption of digital twins and modeling languages in practice, future studies could also investigate the most significant technological challenges and barriers that the industry faces today.} 
\sosym{Regarding informal models our results indicate that they are mostly
used for communication and documentation purposes. They are particularly used during
the conception phase, are mostly drawn in digital tools such as PowerPoint, and
describe the behavior of the CPS. 
Therefore, we see an opportunity in artificial intelligence technology~\cite{KoziolekKoziolek2024_1000175780} to transform these informal models in first versions of formal models in the future.
Moreover, we see major opportunities in the development of intelligent modelling assistents~\cite{mussbacher2020opportunities} for the automotive industry.} 
In general, we believe that the insights gained from our questionnaire survey can help practitioners and researchers engineering automotive \dts more systematically.

\begin{acks}
This work has been partly supported by the German Federal Ministry of Economic Affairs and Climate Action (BMWK, SofDCar – 19S21002L).
The authors of the University of Stuttgart were partly supported by the Deutsche Forschungsgemeinschaft (DFG, German Research Foundation) [grant number 441207927] and by the Ministry of Science, Research and Arts of the Federal State of Baden-Württemberg within the Innovations Campus Future Mobility (ICM).
This work was supported by funding from the pilot program Core Informatics at KIT (KiKIT) of the Helmholtz Association (HGF) and supported by the Deutsche Forschungsgemeinschaft (DFG, German Research Foundation) - SFB 1608 - 501798263.
\end{acks}

\balance
\bibliographystyle{ACM-Reference-Format}
\bibliography{src/bib/andreaswortmann,src/bib/crossdomain,src/bib/references,src/bib/related_work,src/bib/informal-models}

\end{document}